\documentclass{vldb}
\usepackage{graphicx}
\usepackage[export]{adjustbox}
\usepackage{balance}  
\usepackage{enumitem}
\usepackage{hyperref, xcolor}
\hypersetup{colorlinks, citecolor=blue}

\usepackage[T1]{fontenc}
\usepackage[utf8]{inputenc}
\usepackage{authblk}
\usepackage{multirow}
\usepackage{bbding}
\usepackage{pifont}
\usepackage{wasysym}
\usepackage{amssymb}
\usepackage{array,multirow}
\usepackage{cleveref}

\crefformat{section}{\S#2#1#3} 
\crefformat{subsection}{\S#2#1#3}
\crefformat{subsubsection}{\S#2#1#3}

\usepackage{url}
\usepackage{balance}
\usepackage[export]{adjustbox}

\newcommand\longvar[1]{\mathchardef\UrlBreakPenalty=100
	\mathchardef\UrlBigBreakPenalty=100\url{#1}}

\makeatletter
\newcommand{\printfnsymbol}[1]{%
  \textsuperscript{\@fnsymbol{#1}}%
}
\makeatother

\balance
\begin{document}


\title{Blockchain Meets Database: Design and Implementation of a Blockchain Relational Database\thanks{Copyright held by the owner/author(s)}}
\author[1]{\large Senthil Nathan} 
\author[1]{Chander Govindarajan} 
\author[1]{Adarsh Saraf} 
\author[2]{\\Manish Sethi}
\author[1]{Praveen Jayachandran}
\affil[ ]{$^{1}$IBM Research  India,  $^{2}$IBM Industry Platforms USA}
\affil[ ]{\normalsize {\textit{$^{1}$(snatara7, chandg12, adasaraf, praveen.j)@in.ibm.com, $^{2}$manish.sethi1@ibm.com}}}

\maketitle

\begin{abstract}
	In this paper, we design and implement the first-ever decentralized replicated relational
	database with blockchain properties that we term \textit{blockchain relational database}.
	We highlight several similarities between features provided by
	blockchain platforms and a replicated relational database, although they are conceptually different,
	primarily in their trust model. Motivated by this, we leverage the rich features, decades of research
	and optimization, and available tooling in relational
	databases to build a blockchain relational database. 
	We consider a permissioned blockchain model of known, but mutually distrustful organizations
	each operating their own database instance that are replicas of one another.
	The replicas execute transactions independently and engage in decentralized consensus
	to determine the commit order for transactions. We design two approaches, the first where
	the commit order for transactions is agreed upon prior to executing them, and the second where
	transactions are executed without prior knowledge of the commit order while the ordering happens
	in parallel. We leverage serializable snapshot
	isolation (SSI) to guarantee that the replicas across nodes remain consistent and respect the
	ordering determined by consensus, and devise a new variant of SSI based on block height for
	the latter approach.
	We implement our system on PostgreSQL and present detailed performance
	experiments analyzing both approaches.
\end{abstract}

\section{Introduction}
\label{sec:intro}
Blockchain has gained immense popularity over recent years, with its application
being actively explored in several industries. 
At its core, it is an immutable append-only log of cryptographically
signed transactions, that is replicated and managed in a decentralized fashion through
distributed consensus among a set of untrusted parties. 
%
Opinion on the technology has varied widely from being hailed as the
biggest innovation since the Internet to being considered as another database in disguise.
For the first time, we
undertake the challenge of explaining blockchain
technology from the perspective of concepts well known in databases, and highlight the similarities
and differences between the two.

Existing blockchain platforms~\cite{fabric, ethereum, bitcoin, bigchaindb}, in an attempt to build something radically new and transformative,
rebuild a lot of features provided by databases, and treat it as just a store of information. We take a contrarian view in this paper.
By leveraging the rich features and transactional processing capabilities already built into relational
databases over decades of research and development, we demonstrate that we can build a \textit{blockchain
relational database} with all features provided by existing blockchain platforms and with better support for
complex data types, constraints, triggers, complex queries, and provenance queries.
Furthermore, our approach makes building blockchain applications as easy as building database applications;
developers can leverage the rich tooling available for backup,
analytics, and integration with existing enterprise systems. Applications that have a strong compliance and audit
requirements and need for rich analytical queries such as in financial services that are already built atop relational
databases, or ones that are reliant on rich provenance information such as in supply chains are likely
to most benefit from the blockchain relational database proposed in this paper.

With a focus on enterprise blockchain applications, we target a permissioned setup of known but untrusted
organizations each operating their own independent database nodes but connected together to form a blockchain
network. These database nodes need to be replicas of each other but distrust one another. 
The replication of
transaction logs and state among multiple database nodes is possible using
master-slave~\cite{master-slave, master-slave-master-1,  master-slave-1, master-slave-master} and
master-master~\cite{master-slave-master-1, master-master, master-slave-master} protocols. In these protocols,
the master node is trusted as it is
the only node that executes transactions and propagates the final updates to other masters and slaves
either synchronously or asynchronously~\cite{eager-lazy}. For synchronous replication in the Master-Master architecture,
the serializable order is decided using techniques such as 2-phase commit~\cite{2pc}, 3-phase commit~\cite{3pc}, a quorum
protocol~\cite{quorum} to ensure consistency. Asynchronous replication can result in an intermediate inconsistent state
between masters. Various solutions~\cite{lazy-reconcile, lazy-reconcile-1} exist to reconcile inconsistency.
However, in a blockchain network, one node cannot trust the execution
results of another node. 
Hence, all transactions need to be independently executed on all nodes.
Further, such an execution and subsequent commits across multiple untrusted
nodes must result in the same serializable order to ensure
consistency. As a result, existing replication protocols cannot be used for
blockchain networks and hence,
we need a novel protocol with a notion of decentralized trust. Although crash fault-tolerant synchronous
commit protocols~\cite{paxos-commit, pbft-commit} exist, it involves multiple rounds of communications between
database nodes per transaction or a centralized trusted controller. However, applying
synchronous commit per transaction on a blockchain network 
would be very costly
as nodes may be globally distributed. Further, a trusted controller cannot be used.

\begin{table*}[htp]
	{
		\caption{{Similarities between blockchain properties and relational database features.}}
	\label{table:compare-blockchain-rdbms}
	\begin{tabular}{  | p{4.4cm} | p{6.5cm} | p{5.7cm} | }
        \hline
	\textbf{Blockchain Properties} & \textbf{Relational Database Features} & \textbf{Enhancement Needed} \\ \hline
	\textit{Smart contract} & PL/SQL procedure & Deterministic execution \\ \hline
	\textit{Authenticity, non-repudiability} & User management with groups and roles & Crypto-based transaction authentication \\ \hline
	\textit{Access control} & Role \& content-based ACL policies & None \\ \hline
	\textit{Immutable transaction logs} & Transaction logs & Digitally signed transactions \\ \hline
	\textit{Consistent replicated ledger between untrusted nodes} & Master-slave \& master-master replication with trust on transaction ordering and update logs & Decentralized trust and transaction ordering determined by consensus \\ \hline
	\textit{Serializable isolation level} & Strict 2-phase locking, optimistic concurrency control, serializable snapshot isolation & Order must respect block ordering obtained through consensus \\ \hline
	\textit{Async transaction, notification} & LISTEN \& NOTIFY commands & None \\ \hline
	\textit{Provenance queries} & Maintains all versions of a row & Enable query on historical records \\ \hline
  \end{tabular}
}
\end{table*}

\textit{In this paper, we address this key challenge of ensuring that all the untrusted database nodes execute
	transactions independently and commit them
	in the same serializable order asynchronously.}
Towards achieving this, we make two key design choices. First,
we modify the database to separate the concerns of concurrent transaction execution and decentralized ordering
of blocks of transactions. We leverage ordering through consensus only to order blocks of transactions, and not for the
serializable ordering of transactions within a single block.
Second, independently at each node, we leverage serializable snapshot isolation (SSI)~\cite{ssi-first, ssi-second} to
execute transactions concurrently and then serially validate \& commit each transaction in a block.
to obtain a serializable order that will be the same across all untrusted nodes and respect
the ordering of blocks obtained from consensus.
Using these ideas, we present two novel approaches,
the first where block ordering is performed before transaction execution, and the second where transaction
execution happens parallelly without prior knowledge of block ordering, and discuss their trade-offs.
We leverage and \textit{modify serializable snapshot isolation~\cite{ssi-second}
to create a novel variant} which guarantees the same serializable ordering across all nodes.
We implement our ideas on PostgreSQL
and discuss several subtle challenges we had to overcome. 
We make the following contributions in this paper: 
\begin{enumerate}
\item We devise two novel approaches of building a blockchain platform starting from a database, ensuring that
a network of untrusted database nodes can remain consistent. 
\item We devise a new variant of SSI based on block height to permit transaction execution to happen parallelly
with ordering based on consensus. 
\item We implement the system in PostgreSQL that required an addition of only 4000 lines of C code
and present the modified architecture.
\end{enumerate}

The rest of this paper is organized as follows. 
We highlight the properties required in a blockchain platform, motivate how several of these are either partially or fully
available in relational databases, and identify shortcomings that need to be addressed in \S\ref{sec:combined-motivation}.
We discuss our design in \S\ref{sec:design} and describe two approaches to transaction processing leveraging
serializable snapshot isolation. We provide an account of our prototype implementation on PostgreSQL in
\S\ref{sec:implementation}. In \S\ref{sec:experiments}, we present a detailed performance study. We discuss
related work in \S\ref{sec:related}, future work in \S\ref{sec:futurework}, and conclude in \S\ref{sec:conclusion}.

\section{Motivation}
\label{sec:combined-motivation}
Table~\ref{table:compare-blockchain-rdbms} presents a comparison of properties required by a blockchain
platform and features available in relational databases.
We argue that there is much to be gained by leveraging and enhancing the features of relational databases
to build a permissioned blockchain platform, rather than building one from scratch.
This approach helps us leverage the decades of research in
relational databases, stable enterprise-ready code with rich features,
the large community, libraries, tools and integrations with enterprise applications.
Next, we compare blockchains and databases 
across eight features and the enhancements needed.

{\bf (1) Smart contract \& transaction:}
Many blockchain platforms such as
Hyperledger Fabric~\cite{fabric}, Ethereum~\cite{ethereum} and Corda~\cite{corda} support
smart contracts which are functions that manage
state on the blockchain ledger. Transactions are invocations of smart contract functions. Independent
execution of the smart contract functions by different mutually
distrustful peers need to be deterministic.
This is similar to
stored PL/SQL procedures in relational databases. Both accept
input arguments, retrieve data by executing queries, apply
the predefined logic, and write data back to the database. 
However, the execution of a PL/SQL procedure may not be
deterministic when executed independently on multiple nodes 
if utility functions such as \texttt{random} and
\texttt{timestamp} are used. Hence, we need to constrain PL/SQL procedures to ensure that
the execution is deterministic.

{\bf (2) User authenticity \& non-repudiability:} Permissioned blockchain systems
employ public key infrastructure for user management and ensuring authenticity. Participating
users belong to organizations are permissioned to participate in the blockchain network.
Transactions are digitally signed by the invoker and recorded on the blockchain, making them non-repudiable.
Relational databases have sophisticated user management capabilities including support for users, roles,
groups, and various user authentication options. However, submitted and recorded transactions are not signed
by the invoker making them repudiable. Hence, we need invoker's signature on transaction.

{\bf (3) Confidentiality \& access control:} Some permissioned blockchains support the notion of confidentiality, where
smart contracts, transactions and data are only accessible by authorized users. 
In addition, access to invoke functions and to modify data is restricted to specific users.
Relational databases have comprehensive support for access control
on tables, rows, columns and PL/SQL procedures. 
Some relational databases even provide features like content-based access
control, encryption and isolation of data.
Advanced confidentiality features where only a subset of the nodes have access to a particular
data element or smart contract can be implemented, although we don't consider that in this paper.

{\bf (4) Immutability of transactions and ledger state:} The blockchain ledger is an append-only log of 
blocks containing sets of transactions that are chained together by adding the hash of each block
to it's succeeding block. This makes it difficult for an attacker to tamper with a block and avoid
detection. The ledger state can only be modified by invoking smart contract functions recorded as transactions
on the blockchain and are immutable otherwise.
Although a database logs all submitted transactions, executed queries, and the user information, it isn't
possible to detect changes to the log as there are no digital signatures. 

{\bf (5) Consistent replicated ledger across distrustful nodes:} All non-faulty peers in a blockchain must maintain
a consistent replica of 
all transactions and the ledger state. This is ensured by all peers committing transactions 
in the same sequential order as agreed through consensus. Note that consensus is performed on blocks of
transactions rather than on individual transactions to be more performant. 
Databases support replication of
transaction logs and state among nodes using
master-slave~\cite{master-slave, master-slave-master-1,  master-slave-1, master-slave-master} and
master-master~\cite{master-slave-master-1, master-master, master-slave-master} protocols. In these protocols,
the master node is trusted as it is
the only node that executes transactions and propagates the final updates to other masters and slaves.
Hence, we need to
propose a replication protocol with a notion of decentralized trust, while ensuring that the replicas
remain consistent. 

{\bf (6) Serializability isolation, ACID:} Blockchain transactions require serializable isolation, in which
dirty read, non-repeatable read, phantom reads,
and serialization anomalies are not possible. Transactions, when executed in parallel, 
must follow the same serializable order across all nodes.
Further, transactions should be ACID~\cite{acid} compliant.
Serializable isolation can be achieved in databases by employing: (i) strict
2-phase locking~\cite{2pc}, (ii) optimistic concurrency control~\cite{occ}, or
(iii) SSI~\cite{ssi-first, ssi-second}. This needs to be enhanced to follow the block order
as determined through consensus, but can be leveraged to a large extent.

{\bf (7) Asynchronous transactions:} As transaction processing and consensus 
may involve a non-trivial delay, clients submit transactions asynchronously and then leverage
notification mechanisms to learn whether their transaction was successfully committed.
Databases provide support for notification channels and triggers that applications
can leverage.

{\bf (8) Provenance:} The auditable append-only log of transactions in blockchains
can be harnessed as a provenance store, for several use cases including 
supply chain tracking and financial compliance. However, most blockchain platforms today do not support complex queries
on historic data or are not optimized for provenance queries.
In certain relational
databases which support Multi-Version Concurrency Control~\cite{mvcc} such as snapshot isolation~\cite{si},
all versions of a row are maintained and old rows are purged on demand or periodically. However, a SQL
query cannot access the old rows. For provenance queries, we need to disable purging and enable queries on the old rows.

\section{Design of a Blockchain Database}
\label{sec:design}
\begin{figure*}
    \begin{center}
	    \includegraphics[scale=0.54]{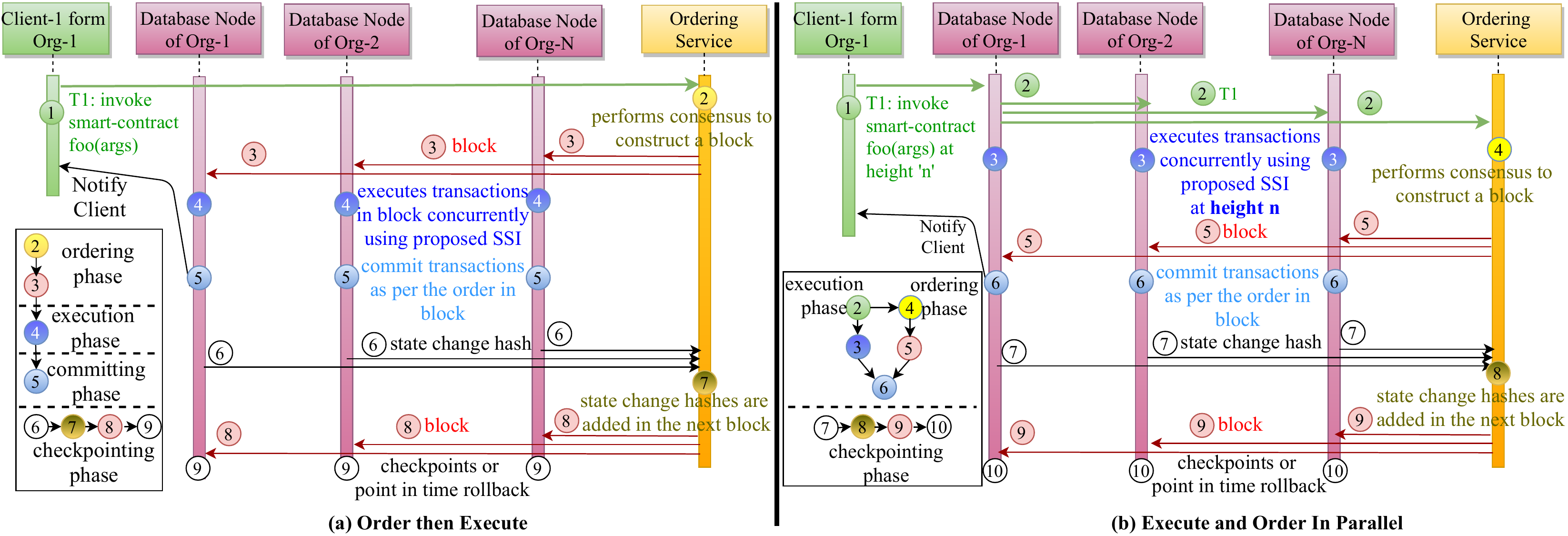}
	    \caption{{Proposed transaction flows---\textit{order-then-execute} and \textit{execute-and-order-in-parallel}}}
	    \label{fig:trans-flow}
 \end{center}
\end{figure*}
We identify two approaches to achieve our goal of building a consistent replicated ledger across
untrusted nodes starting from a relational database.
The first approach, \textit{order-then-execute}, orders all the transactions through a consensus service and then nodes execute
them concurrently, whereas in the second approach, \textit{execute-order-in-parallel}, transaction execution 
happens on nodes without prior knowledge of ordering while block ordering happens parallelly through a consensus service. 
Intuitively, while the first approach is simpler and requires fewer modifications
to the relational database, the second approach has the potential to achieve better performance. We design and
implement both approaches to study their trade-offs.

We first describe the key components of our system in \S\ref{sec:key-components}.
We provide background on Serializable Snapshot Isolation in \S\ref{sec:SSI} and show later in
\S\ref{sec:execute-order} that SSI, if directly applied, does not guarantee serializability and consistency
across nodes for the \textit{execute-order-in-parallel} approach. 
Figure~\ref{fig:trans-flow} juxtaposes the two proposed approaches and we describe them in detail in
\S\ref{sec:order-then-execute} and \S\ref{sec:execute-order}, including a novel SSI based on block height technique.
We discuss security properties of the proposed approaches in \S\ref{sec:security-properties}.
We describe a mechanism for peer node recovery in \S\ref{sec:recovery}
and then discuss how a permissioned blockchain network can be bootstrapped in \S\ref{sec:network-bootstrap}.

\subsection{Key Components}
\label{sec:key-components}
We consider a permissioned network of organizations that are known to one another but may be 
mutually distrustful. The network is private to the participating
organizations and a new organization must be permissioned to join the network. Each organization
may include clients, database peer nodes and ordering service nodes, which we describe below, that together
form the decentralized network.

\textbf{Client:} Each organization has an administrator responsible for onboarding client users
onto the network. The administrator and each client have a digital certificate registered with all the
database peers in the system, which they use to digitally sign and submit transactions on the network
(we describe the network setup process in \S\ref{sec:network-bootstrap}). This helps support
authenticity, non-repudiability and access control properties.
Clients may also listen on a notification channel to receive transaction status.
\par
\textbf{Database Peer Node:} An organization may run one or more database nodes in the network.
All communication to send and receive transactions and blocks happens via a secure communication protocol such as
TLS. Each node also has a cryptographic identity (i.e., public key)
and all communications are signed and authenticated cryptographically. Each database node maintains its own
replica of the ledger as database files, independently executes smart contracts as stored procedures,
and validates and commits blocks of transactions formed by the ordering service.
\par
\textbf{Ordering Service:} Consensus is required to ensure that
the untrusted  database nodes agree on an ordering of blocks of transactions. 
In order to leverage the rich literature on
consensus algorithms with different trust models, such as crash fault tolerant (e.g., Raft~\cite{raft},
Paxos~\cite{paxos}, Zab~\cite{zab}) and byzantine fault tolerant (e.g., PBFT~\cite{pbft},
XFT~\cite{xft}) consensus 
we make the ordering service in our system pluggable and agnostic to the database
implementation.
The ordering service consists
of consensus or orderer nodes, each owned by a different organization. Each orderer node, similar to database
nodes, have their own digital certificate or identity. The output of consensus
yields a block of transactions, which is then atomically broadcast to all the database nodes.
A block consists of (a) a sequence number, (b) a set of transactions, (c) metadata associated with
the consensus protocol, (d) hash of the previous block, (e) hash of the current block, i.e.,
hash (a, b, c, d); and (f) digital signature on the hash of the current block by the orderer node.
\vspace{.5cm}
\subsection{Serializable Snapshot Isolation (SSI)} 
\label{sec:SSI}
Strict 2-phase locking (S2PL), optimistic concurrency control (OCC), and SSI are approaches to achieve serializability.
As SSI offers greater concurrency than
a S2PL and OCC, we have chosen SSI in our design to ensure serializability.
SSI achieves serializability using a modified
snapshot isolation (SI) technique. 
\par
\textbf{Snapshot Isolation and Anomalies.}
In SI, each transaction reads from a consistent snapshot of the database comprising of the last committed values
that existed at the
time the transaction started. Although SI prevents dirty read, non-repeatable
read, and phantom read, it cannot guarantee serializability due to SI anomalies~\cite{si} which violates
consistency (i.e., C in ACID), specifically integrity constraints (refer to \cite{si, ssi-first, ssi-second}
for examples). Hence, Cahill et. al.~\cite{ssi-first} proposed SSI to detect and resolve anomalies automatically to ensure
serializability.
\par
\textbf{Background on SI Anomalies.} To detect SI anomalies, Adya et. al.~\cite{mvsg} proposed Multi-Version
Serialization history Graph (MVSG).
This graph contains a node per transaction,
and an edge from transaction $T_{1}$ to transaction $T_{2}$, if $T_{1}$ must have preceded $T_{2}$ in
the apparent serial order of execution. Three types of dependencies can create these edges:
\begin{itemize}
	\item \textit{rw-dependency}: if $T_{1}$ writes a version of some object, and $T_{2}$ reads the
		previous version of that object, then $T_{1}$ appears to have executed after $T_{2}$, because
		$T_{2}$ did not see the update. To represent this, an edge from $T_{2}$ to $T_{1}$ with a label
		\text{rw} needs to be added. As we will see, these rw-conflicts are central to SSI. Further, a
		dependency can also be caused by predicate reads.
	\item \textit{wr-dependency}: if $T_{1}$ writes a version of an object, and $T_{2}$ reads that version,
		then $T_{1}$ appears to have executed before $T_{2}$ (an edge from $T_{1}$ to $T_{2}$ with a label
		wr).
	\item \textit{ww-dependency}: if $T_{1}$ writes a version of some object, and $T_{2}$ replaces that
		version with the next version, then $T_{1}$ appears to have executed before $T_{2}$ (an edge
		from $T_{1}$ to $T_{2}$ with a label ww).
\end{itemize}
If a cycle is present in the graph, then the execution does not correspond to any serial order, i.e., a SI
anomaly has caused a serializability violation. Otherwise, the serial order can be determined using a
topological sort.
\par
In SI, a \textit{wr} or \textit{ww-dependency} from $T_{1}$ to $T_{2}$ denotes that $T_{1}$ must have committed
before $T_{2}$ began. 
In practice, writes acquire an exclusive lock to prevent \textit{ww-dependency}
between two concurrent transactions. A \textit{wr-dependency} also cannot occur between two concurrent
transactions. In contrast, only \textit{rw-dependency} occurring between concurrent transactions is relevant when studying
serialization anomalies.
\begin{figure}
    \begin{center}
	    \includegraphics[scale=0.8]{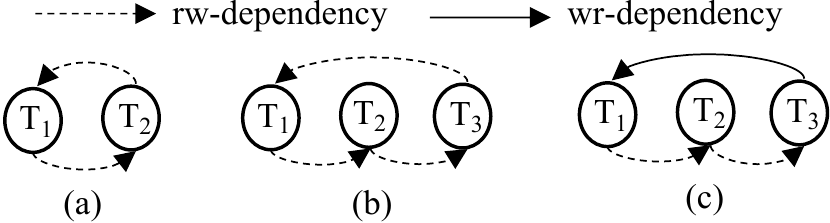}
		\caption{{SI Anomalies between 2 and 3 transactions}}
	    \label{fig:si-anomalies}
 \end{center}
\end{figure}
\par
Adya et. al.~\cite{mvsg} observed that every cycle in a serialization graph contains at least two \textit{rw-dependency} edges.
Fekete et. al.~\cite{mssi} subsequently found that two such edges must be adjacent. Figure~\ref{fig:si-anomalies}(a) shows the only possible
cycle with two transactions, and Figure~\ref{fig:si-anomalies}(b) and (c) show the two possible cycles with three transactions.
If any of the transactions is aborted, a serializable order could be achieved.
\par
\textbf{SSI - Resolving Anomalies.} SSI automatically detects and resolves anomalies. As tracking rw \& wr
dependencies, and subsequently detecting cycles is costly, SSI applies heuristics that are anomaly-safe, but could result in
false positives. They are:
\par
\textit{(1) abort immediately}: Cahill et. al.~\cite{ssi-first} used two flags per transaction T: (a) \textit{inConflict}---set
		when there is a \textit{rw-dependency}
		from a concurrent transaction to T; (b) \textit{outConflict}---set when there is a \textit{rw-dependency} from T to
		a concurrent transaction. As soon as both of these flags are set for T, which could lead to an SI anomaly,
		T is aborted.
\par
\textit{(2) abort during commit}: Ports et. al.~\cite{ssi-second} maintained two lists per transaction T: (a)
		\textit{inConflictList}---maintains a list of
		concurrent transactions from which there is a \textit{rw-dependency} to T.  (b) \textit{outConflictList}---maintains a list
		of concurrent transactions to which T has a \textit{rw-dependency}. The transactions present in inConflictList of T
		are called \texttt{nearConflicts}. The transactions present in the inConflictList of each nearConflict are
		called \texttt{farConflicts}. For e.g., in Figure~\ref{fig:si-anomalies}(b), for $T_{2}$, $T_{1}$ is a nearConflict
		and $T_{3}$ is a farConflict. Recall that \textit{rw-dependency} occurs only between concurrently executing transactions
		(such as in Figures~\ref{fig:si-anomalies}(a) and (b)). For each pair of nearConflict and
		farConflict, if both transactions are not yet committed, then this heuristic aborts the nearConflict so that an immediate retry
		can succeed. In contrast, a \textit{wr-dependency} only occurs between a committed and a just commenced transaction
		(Figure~\ref{fig:si-anomalies}(c)). In this case, only if $T_{3}$ has committed first, its nearConflict $T_{2}$ is aborted.
		Otherwise, no transactions are aborted.
		In other words, while the heuristic does not track \textit{wr-dependency}, it accounts for its possibility and aborts
		a transaction whose outConflict has committed.

		In the next two sections, we describe the transaction flows for 
(1) Order then Execute;
and (2) Execute and Order in Parallel as shown in Figure~\ref{fig:trans-flow}. In both flows, transactions
are committed asynchronously across nodes.
\subsection{Order then Execute}
\label{sec:order-then-execute}
A transaction submitted by a client in the \textit{order-then-execute} approach comprises of (a) a unique identifier,
(b) the \textit{username} of the client, (c) the PL/SQL procedure execution command with the
name of the procedure and arguments, and (d) a digital signature on the \texttt{hash(a, b, c)} using the
client's private key. The transaction flow consists of four pipelined phases: ordering, execution, committing,
and checkpointing, wherein a submitted transaction needs to first be ordered, then executed, and
finally committed before recording a checkpoint. 
\par
\subsubsection{Ordering phase}
\label{sec:ordering-phase-oe}
Clients submit transactions directly to any one of the ordering service nodes
(it is possible for a peer to act as a proxy for the client to submit transactions to the ordering service,
as long as the client trusts the peer to not drop its transactions). 
On a periodic timeout (say every 1 second), the ordering service nodes initiate a consensus protocol among
themselves to construct a block of transactions. After constructing a block, the ordering service delivers
it to all nodes in the network via atomic broadcast. In Figure~\ref{fig:trans-flow}(a),
steps 2 and 3 denote this phase. 
\par
\subsubsection{Execution phase} On receiving a block of transactions,
\label{sec:execution-phase-oe}
each database node verifies whether the received block is in sequence and sent by the ordering service.
On successful verification, the node appends the block to a block store which is maintained in
the local file system. In parallel, the node retrieves unprocessed blocks one at a time, in the order
of their block sequence number, from the block store and performs the following four operations:
\begin{enumerate}
	\item The database node assigns a thread per transaction (provided that the identifier present in a transaction's
		field is unique) to authenticate and execute it. In an append-only \textit{ledger table}, it records each transaction
		in a block. This \textit{ledger table} is used for
		recovery as explained in \S\ref{sec:recovery} and also for provenance queries as demonstrated in
		\S\ref{sec:psql-new-components}.
	\item Each thread retrieves the public key associated with the \textit{username}
		in the transaction, to verify the user's digital signature. On successful
		authentication, the \textit{username} is
		set for the thread's session which is needed for access control. We leverage the database's
		access control capabilities without modification.
	\item Each thread executes the PL/SQL procedure with the passed arguments as per the
		transaction. To ensure that on all nodes the transactions are executed on the same
		committed state, all valid transactions in a block are executed concurrently using the
		\textit{abort during commit} SSI variant. This helps in ensuring that the set of transactions
		marked to be committed are the same and that they follow the same serializable
		order across all nodes. 
	\item Once a thread completes execution of a transaction, the transaction would be ready
		to either commit or abort (as per the procedure's execution logic), but waits without
		proceeding. This is because it could result in a different commit order in different
		nodes, if the execution completion order is different (which could result in different
		aborts by SSI in each node).
\end{enumerate}
Only after committing a block of transactions, the execution phase will process the next block.
In Figure~\ref{fig:trans-flow}(a), step 4 denotes the execution phase.
\par
\subsubsection{Committing Phase} To ensure that the commit order is the same on all nodes, the order in which
\label{sec:committing-phase-oe}
the transactions get committed is the order in which the transactions appear in the block.
Only when all valid transactions are executed and ready to be either committed or aborted, the node
serially notifies one thread at a time to proceed further. Every transaction applies the
\textit{abort during commit} approach to determine whether to commit, and only then does the next transaction enter the
commit stage. While it is possible to apply SSI for all transactions at once to obtain a serializable order,
we defer such optimizations for simplicity. The node records the transaction status in the \textit{ledger table}
and emits an event via
a notification channel to inform the client. In Figure~\ref{fig:trans-flow}(a), step 5 denotes the committing phase.

There is one additional challenge. In SSI, \textit{ww-dependency} is handled using an exclusive lock.
For example, if $T_{1}$ and $T_{2}$ are competing to write an object, only one transaction can
acquire a lock (say $T_{1}$) and the other (say $T_{2}$) must wait for the lock to be released which can happen
only after the commit/abort of $T_{1}$. However, in our system, to ensure consistency across all nodes, the committing phase
cannot start unless all transactions complete execution. So, we cannot use locks for \textit{ww-dependency}. However,
as the ordering between transactions is determined by consensus and fixed across all nodes, we leverage this to
permit both transactions to write to different copies of the object, but subsequently commit only the first as determined
by the ordering. We show how such an implementation is possible in \S\ref{sec:implementation}.

It is noteworthy that certain existing blockchain platforms such as Ethereum also follow an
\textit{order-then-execute} approach. However, they execute transactions sequentially in the order they appear in the block
to ensure serializability and consistency across nodes, but this affects performance. In contrast, leveraging SSI to execute
transactions concurrently and then sequentially issuing committing as per ordering from consensus, enables us to
optimize performance.
\par
\subsubsection{Checkpointing Phase}
\label{sec:checkpointing-phase-oe}
Once all transactions in a block are processed, each node computes the hash of the write set,
which is the union of all changes made to the database by the block, and submits it to the ordering service as a proof of
execution and commit. When a node receives subsequent blocks, it would receive the hash of write set
computed by other nodes. The hash computed by all non-faulty nodes would be the same and the node then proceeds to record a checkpoint
(note that this is different from the Write-Ahead Logging checkpointing). It is
not necessary to record a checkpoint every block. Instead, the hash of write sets can be computed for a
preconfigured number of blocks and then sent to the ordering service.
In Figure~\ref{fig:trans-flow}(a), steps 6, 7, and 8 denote the checkpointing phase.
\subsection{Execute and Order in Parallel}
\label{sec:execute-order}
A transaction submitted by a client in the \textit{execute-order-in-parallel} approach comprises of
(a) the \textit{username} of the client, (b) the PL/SQL procedure execution command with the
name of the procedure and arguments, (c) a block number, (d) a unique identifier which is computed as
\texttt{hash(a, b, c)}, and (e) a digital signature on the \texttt{hash(a, b, c, d)} using the
client's private key. The transaction flow consists of four phases: execution, ordering, committing, and
checkpointing phase.
A submitted transaction is executed by the database nodes and in parallel ordered by the ordering nodes
and placed in a block. This is followed by the commit and checkpoint phases. We describe each phase in detail below.
\par
\subsubsection{Execution Phase} 
\begin{figure}[!t]
    \begin{center}
     \includegraphics[scale=0.5]{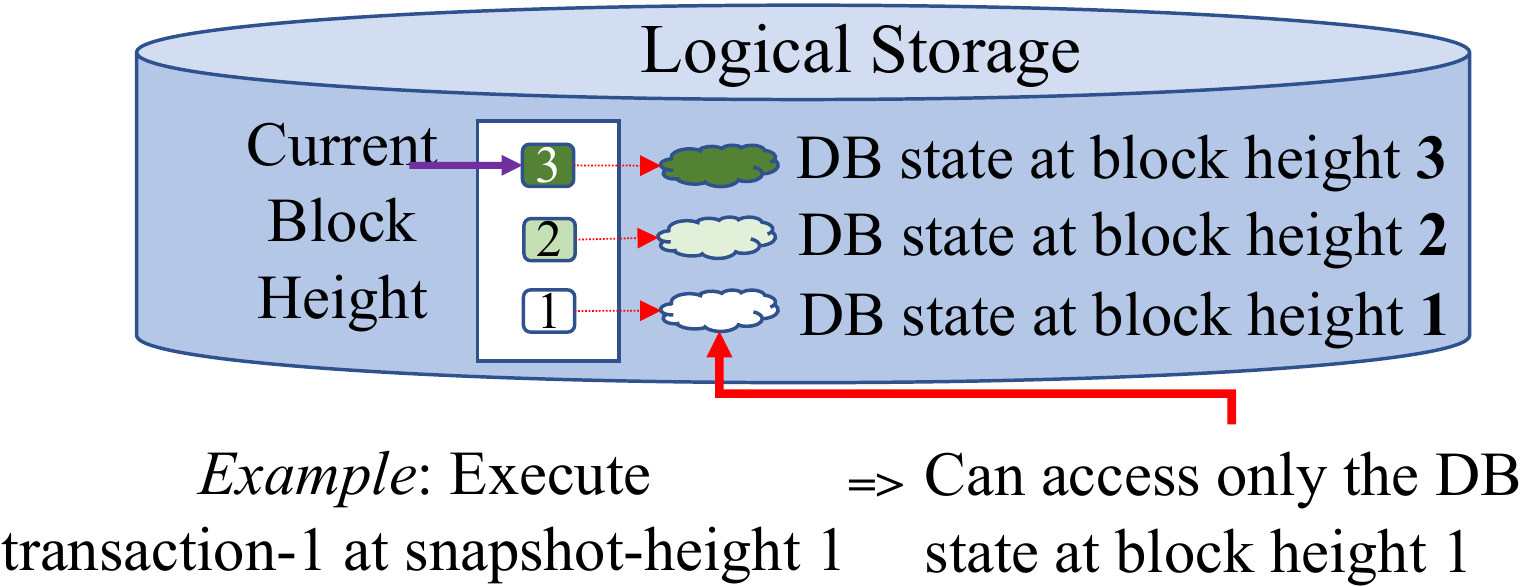}
  \caption{{Logical Database Storage to Enable Snapshot Isolation Based on Block Height.}}
    \label{fig:si-bh}
 \end{center}
\end{figure}

Clients submit transactions directly to one of the database nodes. When a node
\label{sec:execution-phase-eo}
receives a transaction, it assigns a thread to authenticate, forward and execute the transaction.
On successful authentication (same as in the \textit{order-then-execute} approach), the transaction
is forwarded to other database nodes and the ordering service in the background. The four steps
described in \S\ref{sec:execution-phase-oe} for the execution phase applies for the \textit{execute-order-in-parallel}
approach as well, except for the default SSI and the logic to handle duplicate transaction identifiers.
This is shown in steps 1-3 in Figure~\ref{fig:trans-flow}(b).
Unlike the \textit{order-then-execute} approach where execution of transactions happen \textit{after} ordering and on the
committed state from the previous block, in the \textit{execute-order-in-parallel} approach we endeavor
to execute and order in parallel.
To ensure that the transaction is executing on the same committed
data on all nodes, the submitted transaction includes the block height on which it should be executed (the
client can obtain this from the peer it is connected with).
We propose SSI based on block height as depicted in Figure~\ref{fig:si-bh}, as
the default SSI cannot ensure that transactions are executed on the same committed state on all nodes
as each node can be at a different block height depending on its processing speed.
\par
\underline{\textbf{SSI Based on Block Height}}: In this isolation level, each row of a table contains a \texttt{creator}
block number and \texttt{deleter} block number which denote the block that created and deleted this row, respectively
(irrespective of the transaction within the block that created it). Note that, \texttt{creator} and \texttt{deleter} block
number are part of the implementation for the \textit{order-then-execute} approach as well, but are used only for provenance queries
and are not required for SSI.
During a transaction execution, based on the block number specified, the transaction can view the database state
comprising of all commits up to this block height as shown in Figure~\ref{fig:si-bh}. We refer to this specified block number
as the transaction's \texttt{snapshot-height}. In other words,
a transaction sees all rows with \texttt{creator} lesser than or equal to its \texttt{snapshot-height}. For all such rows, the
\texttt{deleter} should be either empty or greater than the \texttt{snapshot-height}. For this SSI to work, we need to maintain all versions of
a row and every update should be a flagging of the old row and an insertion of the new row. Every delete should be
a marking of \texttt{deleter}.
This facilitates the transaction to be executed on the same committed data on all nodes.
\par
However, the current block (\texttt{current-block}) processed for commit by a node might be either lower or
higher than the specified \texttt{snapshot-height} of a transaction. If the \texttt{current} \texttt{-block} is lower, the
transaction would start executing once the node completes processing all blocks and transactions up to the
specified \texttt{snapshot-height}. If the \texttt{current-block}
is higher, the transaction would be executed immediately, but the serializability requirement could be violated 
because of a phantom or stale data read, which needs to be detected and handled. For
example, assume that a transaction is updating all rows in a table which satisfy a given predicate mentioned in
a \texttt{WHERE clause}. There is a possibility that a row that satisfies the predicate was committed
by a block with a number which is greater than
the specified \texttt{snapshot-height} and lesser than or equal to \texttt{current-block}.
In this scenario, a phantom read~\cite{si} has occurred that violated the serializability. Similarly, a transaction
can read a row from a snapshot as of \texttt{snapshot-height}, but if that row was either updated or deleted
by a subsequent block it would result in a stale read for this transaction.
\par
In order to ensure serializability, the proposed SSI approach detects such phantom reads and stale data reads
to abort the corresponding transaction. To detect and abort such a transaction, the proposed approach applies
row visibility logic on the committed state:
\begin{enumerate} 
	\item when a row with \texttt{creator} greater than the specified \texttt{snapshot-height} satisfies
		the given predicate, abort the transaction provided that the \texttt{deleter} is empty (handles phantom read);
	\item when a row with \texttt{creator} lesser than \texttt{snapshot-height} satisfies the given predicate,
		abort the transaction provided that the \texttt{deleter} is non-empty and greater than \texttt{snapshot-height} 
		(handles stale read).
\end{enumerate}
Further, concurrent transactions which are going to be committed during or after \texttt{current-block}
can also create a phantom read or a stale read problem. Such cases are tracked by our
proposed SSI as described in the committing phase.
\par
\subsubsection{Ordering Phase} Database nodes submit transactions to the ordering service unlike the
\label{sec:ordering-phase-eo}
\textit{order-then-execute} approach. Note that the transactions are being executed in the database nodes while they are
being ordered by the ordering service. The rest of the logic is same as explained in
\S\ref{sec:ordering-phase-oe} for the \textit{order-then-execute} approach.
In Figure~\ref{fig:trans-flow}(b), the steps 4 and 5 denote the ordering phase.
\par
\subsubsection {Committing Phase} Similar to the \textit{order-then-execute} approach, an important pre-condition 
\label{sec:committing-phase-eo}
for entering commit phase is that all transactions in a block must have completed its execution
and waiting to proceed with commit/abort. However, there are two key differences compared to the commit phase
of \textit{order-then-execute}.
First, after receiving a block, if all transactions are not running (this occurs if the node has not received
communication of a transaction from another peer due to maliciousness or network delay), the committer starts executing
all missing transactions and waits for their completion before proceeding to the committing phase.
Second, unlike the previous approach, it is possible for concurrent transactions to be executing
at different snapshot heights (as specified by the respective clients). Further, transactions that are concurrent on
one node, may not be concurrent on another, but we need the set of transactions that are decided to be committed to be the same
on all nodes. As a result, we don't support blind updates such as \texttt{UPDATE table SET column = value;} which might
result in a lock for \textit{ww-dependency} only on a subset of nodes. Note, ww-dependencies are handled in the same way as
described in \S\ref{sec:committing-phase-oe}.
Further, instead of employing the
\textit{abort during commit} variant of SSI, we propose a \textit{block-aware abort during commit} variant of SSI
as described next.
\par
{

	\begin{table}
		\small{
	\centering
	\caption{\small{Abort rule for our proposed SSI variants when Transaction T is committing.}}
	\label{table:abort-rule}
	\begin{tabular}{ c | c | c | c }
        \hline
	{nearConflict} & {farConflict} & to commit first & {Abort} \\ \cline{0-1}
		\multicolumn{2}{c|}{same block} & among conflicts&\\ \hline
	\CheckmarkBold & \CheckmarkBold & nearConflict & farConflict \\ \hline
	\CheckmarkBold & \CheckmarkBold & farConflict & nearConflict \\ \hline
	\CheckmarkBold & \XSolidBrush & nearConflict & farConflict \\ \hline
	\XSolidBrush & \CheckmarkBold & farConflict & \multirow{3}{*}{nearConflict} \\ \cline{0-2}
	\XSolidBrush & \XSolidBrush & - &  \\ \cline{0-2}
	\XSolidBrush & none & - &  \\ \hline

  \end{tabular}
	\vspace{-.2cm}
}
\end{table}
}

\underline{\textbf{{SSI variant---block-aware abort during commit}.}} Table~\ref{table:abort-rule} presents
the abort rules for the proposed SSI variant. In addition to
\texttt{nearConflict} and \texttt{far-} \texttt{Conflict} used in \textit{abort during commit},
our variant considers two additional parameters: 
\begin{enumerate} 
	\item whether the \texttt{nearConflict} and \texttt{farConflict} are in the same block.
	\item if they are in the same block, which among them is earlier as per the ordering. 
\end{enumerate}
	When either or both the conflicts are in the same block
as transaction T, it is straightforward to abort the one that comes later in the ordering, and is deterministic on all nodes.
\par
The tricky case is when neither of the conflicts are in the same block. In this case, we abort the \textit{nearConflict} 
transaction. Note, the nearConflict
is not in the same block but executes concurrently with $T$---this means that the \texttt{snapshot}
\texttt{-height} specified for both
could be lesser than or equal to the current block height at the node. With no synchronization on
transaction executions between nodes, it is possible for an SI anomaly to occur only at a subset of nodes.
To ensure consistency between nodes, we need to ensure that the same set of transactions are aborted on
all nodes. Let us consider possible scenarios in other nodes (say $T_{2}$ is the nearConflict transaction):
\begin{enumerate}
	\item If $T_{2}$ is concurrent with $T$ and an anomaly structure is detected, then $T_{2}$ is
		aborted as per our heuristic.
	\item If $T$ commits before $T_{2}$ starts execution, then $T_{2}$ being a nearConflict for $T$
		read a stale value and would be aborted as discussed earlier.
	\item If $T_{2}$ is concurrent with $T$, but $T$ is committed first, then this is a case of
		a \textit{rw-dependency} where the outConflict has already committed leading to an
		anomaly structure (similar to Figure~\ref{fig:si-anomalies}(c)) and $T_{2}$
		will be aborted in this case as well.
\end{enumerate}
Even if there is no farConflict, the
nearConflict would get aborted (if it not in same block as T) as it could result in a stale read only
at a subset of nodes. Whereas, in other nodes, it might result in a rw-dependency without an anomaly
structure.
Hence, we must abort the nearConflict irrespective of the presence of a farConflict, whenever the nearConflict 
is not in the same block.
\par
\par
The unique identifier used for a transaction must be the hash of (a) the \textit{username} of the client,
(b) the PL/SQL procedure execution command with the name of the procedure and arguments, and (c) a block number specified
for the SSI by the client. The reason is that if two different transactions are submitted to two different nodes
with the same unique identifier, whichever transaction executes first on a given node is the one that would succeed,
whereas the other would fail due to the duplicate identifier. As this can result in an inconsistent state across nodes,
the unique identifier is composed of the three fields in the transaction to ensure that no two different
transactions have the same identifier.
\subsubsection{Checkpointing Phase} This phase is same as the one explained in \S\ref{sec:checkpointing-phase-oe}.
\label{sec:checkpointing-phase-eo}
\subsection{Discussion on Security Properties} \label{sec:security-properties}
\par
\textbf{(1) Submission of invalid transactions.}
A malicious client can potentially submit a large number of invalid transactions (e.g., ones that
will eventually fail due to stale reads, try to perform operations they do not have access for)
in an attempt to launch a denial of service attack. This can be thwarted in one of two ways. First, similar
to permissionless blockchain networks, a transaction fee could be levied for each transaction using a native currency
(it is possible to implement a native currency in our system if desired). Second, by monitoring clients and their
behavior on the network, it is possible to exclude them from participation. We leave such enhancements for future
work as it does not affect the core design of our system.
\par
\textbf{(2) Obscuration of valid transactions.}
In \textit{order-execute} approach, when a malicious orderer node receives a transaction from the user, it might
not include the transaction in a block. Similarly, in \textit{execute-order-in-parallel} approach, when a malicious
database node receives a transaction from the user, it might not forward the transaction to other database and
orderer nodes. In both scenarios, at the client side, the transaction request would timeout. The client
can then submit the same transaction to some other orderer or database node depending on the approach. 
Note that even if the client side timeout was a false alarm (i.e.,
the transaction is forwarded, included in a block and executed), the resubmission of the same transaction
does not affect the data consistency as all nodes check for the unique identifier included in the transaction
before execution.
\par
In \textit{execute-order-in-parallel} approach, if the database node forwards the transaction to an orderer node but
not to other database nodes, eventually the request would be added into a block and delivered to all database nodes.
The default committing phase described in \S\ref{sec:execute-order} would take care of executing missing transactions
while processing the block.
\par
\textbf{(3) Withholding of a block or transaction commit.}
In both approaches, when a malicious database node receives a block, it might skip committing the block or a particular
transaction in that block. In such a scenario, the hash computed during the checkpointing phase would differ from other
nodes, and it would become evident during the checkpointing process that the malicious node did not commit the block
correctly. As other organizations can detect such maliciousness, there is no incentive for the malicious
organization to engage in such a scenario. Since, database nodes validate and commit blocks independently of one another,
a malicious database node cannot hamper the liveness of the network.
\par
\textbf{(4) Byzantine ordering nodes.}
A malicious ordering node could send an incorrect block to database nodes connected to it. If an organization's
database node trusts the ordering node it operates, but the ordering node is malicious, then that database node would
also become faulty. This scenario would then be similar to the previous scenario of the database node being malicious.
If a database node does not trust the ordering node, it should obtain blocks from $2f+1$ ordering nodes (assuming the
consensus algorithm tolerates $f$ failures), to ensure it obtains the correct block after consensus.
\par
\textbf{(5) Tampering of user data.}
In both approaches, the user can submit a read query to a single database node to fetch the stored data. A malicious
database node can tamper the data and return an incorrect result to the user. The following are two ways to detect such malicious
behavior. The user can submit the query to multiple database nodes and verify whether the results are the same. Otherwise, any of the
existing query authentication~\cite{qa1, qa3, qa2, qa5, qa4} methods can be used to verify the integrity of query results. Further, when a stored
data is tampered with, it would eventually be identified through the checkpointing phase.
\par
\textbf{(6) Tampering of blockchain.}
Each database node stores all blocks in a separate store called the \textit{block store}.
If any malicious
database node tampers its block store, it will not affect the other replicas maintained at other organizations'
node. In order to tamper the block store and not be detected, the database node would need the private cryptographic key of the orderer
node as well as the client who submitted the transaction to forge their signatures. Even if the malicious node 
achieves this, if the majority of the nodes are non-malicious, honest organizations could
prove that the malicious organization has tampered with the block store by comparing the replicated chain when a conflict
occurs. Only if 51\% of the nodes are malicious, a blockchain can be successfully tampered.
\subsection{Recovery After a Failure}
\label{sec:recovery}
A blockchain network is required to tolerate node failures 
and we need to ensure that
a failed node recovers to a consistent state when it restarts. In both \textit{order-then-execute} and
\textit{execute-order-in-parallel} approaches, the block processing stage has the following two common
operations per block:
(1) atomically store all transactions information in the \textit{ledger table} along with the
block number; (2) atomically store all transactions' status (i.e., commit/abort) in the \textit{ledger table}.
Note, only after all transactions get written to write-ahead-logging and the default transaction logs, the
step 2 gets executed. A node can fail during any of the above two stages. When the node comes
back online, it would execute the following operations to recover the node to a consistent state:
\begin{enumerate}
	\item Retrieves the last processed block number from the \textit{ledger table} and checks whether all
		transactions in the \textit{ledger table} have a status. Note, as we store all transactions'
		status atomically, either all transactions must have a status or none.
	\item If a status is found in the \textit{ledger table} for transactions, it means that the block was
		committed successfully and
		no recovery is needed. Note, if a node fails after committing a transaction, during restart,
		the default recovery mechanism that uses write-ahead logging (WAL) would take care of disk consistency.
	\item If a status is not found in the \textit{ledger table} for transactions, a node might have
		(a) either failed after committing
		all those transactions but before updating the \textit{ledger table} with the status, or (b) failed before
		committing all or some those transactions. If a transaction was successfully committed, the relational database
		would have recorded the status on the default transaction logs. The node first checks the
		status of each transaction by reading the default transaction logs. If a status is present for all
		transactions, the node would update the \textit{ledger table} with the status (i.e., case (a)).
	\item If a status is not present in the transaction log for all or some transactions (i.e., case (b)), the node needs to
		rollback all other committed transactions in the same block. Then the node can start re-executing all
		transactions as described in \S\ref{sec:execution-phase-oe} and \S\ref{sec:execution-phase-eo},
		commit them, and record the status in
		the \textit{ledger table}. The rollback of
		committed transactions is required as we need to execute all transactions in a block parallelly
		using SSI at the same time to get a consistent result with other nodes as well.
\end{enumerate}
By the time a node restarts and recovers after a failure, a few more blocks may have been added in the network. When the node
receives a block from the ordering service, it would notice a gap in the sequence number due to missing blocks.
To ensure consistency, the node then retrieves any missing blocks, processes and commits them one by one. 
We observe that it is possible to do a single atomic commit for all transactions in a block after validating them serially.
This can simplify the recovery process described above, but we have omitted describing this in the interest of simplicity.
\subsection{Network Bootstrapping}
\label{sec:network-bootstrap}
To bootstrap a permissioned network, first, we need to bring up the database and orderer nodes each with their
respective admin users and secure communication via TLS between them.
We need to then create
users \& roles for each organization, deploy PL/SQL procedures as smart contracts, and define access control policies.
\par
\textbf{Setting up database nodes and ordering service.} 
At network startup, each organization shares data consisting of TLS certificates,
domain names of database and orderer nodes, and credentials of admin users (i.e., public keys) with all other organizations. 
Each organization then starts its database and orderer nodes with the credentials of all organizations' admin users.
\par
\textbf{Support for blockchain and non-blockchain schema.}
As a part of database node startup, the node creates
a default database with two schemas named \textit{blockchain} and \textit{non-blockchain} (the latter being optional
for any private data held by the organization).
Any transaction submitted to a blockchain schema is treated as a blockchain transaction, which must follow the
transaction flow described in \S\ref{sec:order-then-execute} or \S\ref{sec:execute-order}. In the blockchain schema, both users and admins can
execute only PL/SQL procedures and individual \texttt{SELECT} statements.
Note, individual \texttt{SELECT} statements would not be recorded on the blockchain and can only be a read-only
operation. Other individual DML \& DDL statements are not allowed to be executed, but only through PL/SQL procedures.
On a non-blockchain schema, transactions are executed using
the default single node transaction flow implemented in a relational database. The non-blockchain schema in a
database node is accessible only by the organization which owns that node. 
of other organizations can query or submit transactions only on the blockchain schema of the database node and not on the
non-blockchain schema. 
Users of an organization can execute reports or analytical queries combining the
blockchain and non-blockchain schema on the node owned by their organization.
\par
\textbf{Setting up of PL/SQL procedures.} To facilitate deployment of smart contracts, each node exposes
the following four system smart contracts. These contracts are created in the blockchain schema during the database node startup.
These system smart contracts can only be invoked by organization admins, are considered blockchain transactions and follow the 
transaction flow described earlier.
\begin{enumerate}
	\item \texttt{create\_deployTx()} creates, replaces or drops a smart contract. 
		The argument to this procedure can be \texttt{CREATE FUNCTION} statement (to create a new smart-contract),
		\texttt{CREATE OR REPLACE FUNCTION} (to update an existing smart-contract), or \texttt{DROP FUNCTION}
		(to delete an existing smart-contract). On a successful execution, this procedure creates a record in the
		deployment table with the passed argument but does not yet execute it to \texttt{CREATE/REPLACE/DROP}.
	\item \texttt{submit\_deployTx()} executes the SQL statement present in the deployment table after verifying that 
		an admin from each organization has
		approved the deployment transaction. If not all organizations have approved, the invocation returns an error.
		If a smart contract is updated, any uncommitted transactions that executed on an older version of the contract are aborted.
	\item \texttt{approve\_deployTx()} approves a deployment transaction by adding a digital signature provided by the
		organization's admin.
	\item \texttt{reject\_deployTx()} rejects a deployment transaction by adding a digital signature provided by the
		organization's admin and a reason for the rejection.
	\item \texttt{comment\_deployTx()}: adds a comment to a deployment transaction. When an admin wants to suggest
		a change to a smart contract present in a deployment transaction, this system contract can be used.
\end{enumerate}
Further, each node creates three more system smart contracts to create, delete, and update users with cyrptographic
credentials. 
As an invocation of all system smart contracts are recorded as blockchain transactions, the network holds an
immutable transaction log and a history of approvals from organizations before deploying smart contracts.
For access control policies, the network cannot rely on the system contract as it is unware of tables to be created
and privileges for each user. Hence, access control policies need to be embedded within a smart contract itself.
Once user's smart contracts are created successfully using the above system contracts, applications can start
invoking them.

\section{IMPLEMENTATION USING POSTGRESQL}
\label{sec:implementation}
\begin{figure*}
    \begin{center}
	    \includegraphics[scale=0.4]{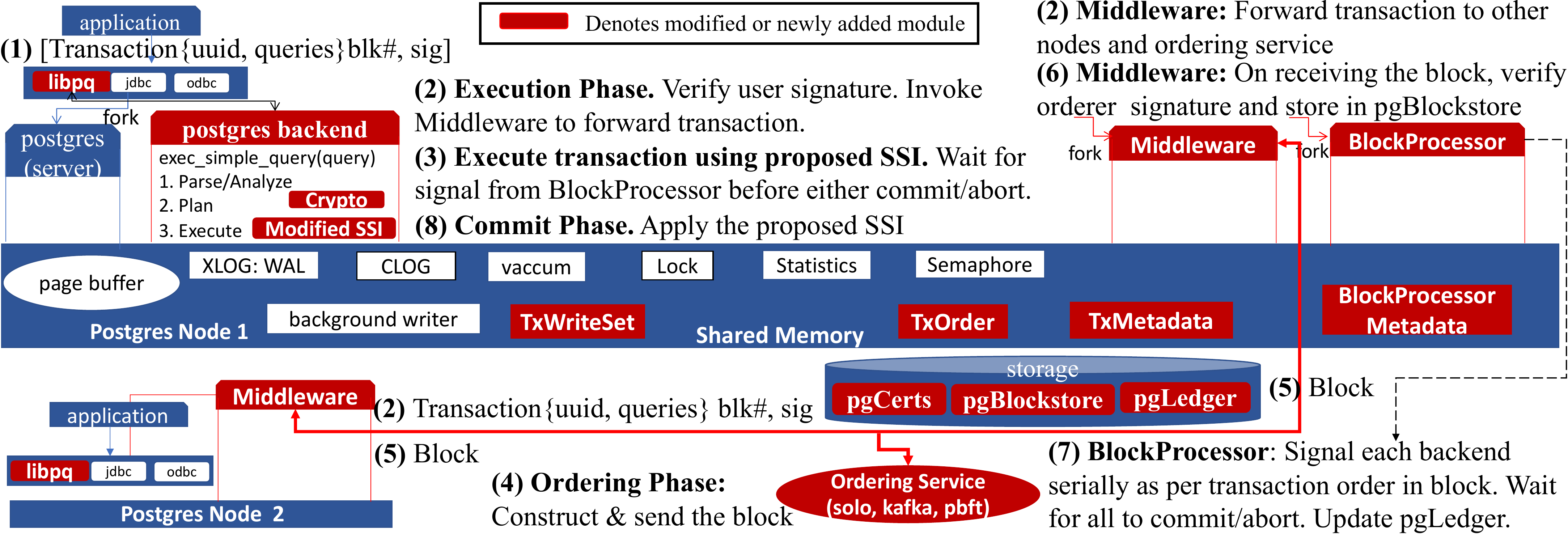}
  \caption{\small{Transaction Flow in PostgreSQL for \textit{execute-order-in-parallel}}}
	    \label{fig:trans-flow-execute-and-order-postgres}
 \end{center}
\end{figure*}

PostgreSQL~\cite{psql} is the first open source database to implement the
\textit{abort during commit} SSI variant~\cite{ssi-second}.
Further, it maintains all rows even after an update or delete and is highly modular and extensible.
For these reasons, we chose to modify PostgreSQL to build a blockchain relational database, rather
than implementing one from scratch. The modifications only amounted to adding around 4000 lines of
C code to implement both approaches. We present relevant background on PostgreSQL in \S\ref{sec:psql-background}. Then, in
\S\ref{sec:psql-new-components} and \S\ref{sec:psql-modified-components},
we describe the components we added and modified in PostgreSQL respectively.
In \S\ref{sec:psql-ordering-service}, we present implementation details of a crash fault-tolerant
ordering service, and describe the transaction flows in \S\ref{sec:psql-transaction-flow}.


\subsection{PostgreSQL Background} 
\label{sec:psql-background}
PostgreSQL supports three isolation levels---read committed, repeatable read
(i.e., SI), and serializable (i.e., SI with detection and mitigation of
anomalies). A snapshot comprises of a set of transaction IDs, which were committed
as of the start of this transaction, whose effects are visible.
Each row has two additional elements in the header, namely \textit{xmin} and \textit{xmax},
which are the IDs of the transactions that created
and deleted the row, respectively. Note, every update to a row is a delete followed by an
insert (both in the table and index). Deleted rows are flagged by setting xmax instead of being actually deleted.
In other words, PostgreSQL maintains all versions of a row
unlike other implementations such as Oracle that update rows in-place and keep a
rollback log. This is ideal for our goal of building a blockchain that maintains all versions
of data. For each row, a snapshot checks \textit{xmin} and \textit{xmax} to see
which of these transactions' ID are included in the snapshot to determine row visibility.
\par
Clients connect to the PostgreSQL server, \texttt{postgres}, using an
application interface such as libpq~\cite{libpq}. A backend
process~\cite{backends} is then assigned to each client connection to execute
queries.
PostgreSQL supports background workers~\cite{bgworkers} (additional processes other than
backend) to perform other activities such as logical replication and statistics collection.
Note, the \texttt{postgres} server, backends, and background workers coordinate and share data through
shared memory regions. Both the shared memory data structures and background workers are easily
extensible.
\subsection{New Components Introduced} 
\label{sec:psql-new-components}
\par
\textbf{Communication Middleware \& Block Processor.}
We introduced two new background workers: (1) communication middleware to communicate with
other nodes and orderer, to transfer and receive transactions/blocks. The received blocks
are stored in an append-only file named \textit{pgBlockstore}. Further, the
middleware is also responsible for starting a transaction using libpq in the 
\textit{order-then-execute} flow, and for starting any missing transactions in the \textit{execute-order-in-parallel}
flow;  
(2) block processor to process a block. It executes
the commit phase as described in \S\ref{sec:committing-phase-oe} and
\S\ref{sec:committing-phase-eo}.

\begin{table*}
	{
		\caption{{Example provenance queries to audit user transactions.}}
	\label{table:audit}
	\begin{tabular}{  | p{8cm} | p{8.8cm} | }
       	\hline
	\textbf{\centering{Audit Scenarios}} & \textbf{\centering{Queries}} \\ \hline
	Get all invoice details from table \textit{invoices} which were updated by a supplier \textit{S} between blocks \textit{100} and \textit{300}
	&
	\texttt{SELECT} invoices.* \texttt{FROM} invoices, pgLedger
	\texttt{WHERE} pgLedger.blockNumber \texttt{BETWEEN} 100 \texttt{AND}
	300 \texttt{AND} pgLedger.user = S \texttt{AND} invoices.xmax =
	pgLedger.txid \texttt{AND} invoices.deleter \texttt{IS NULL}; \\ \hline
	Get all historical details of an invoice from table \texttt{invoices} whose \textit{invoiceID} (primary key) is \textit{k} and was updated by either supplier \textit{S} or manufacturer \textit{M} in the last 24 hours
	&
	\texttt{SELECT} invoices.* \texttt{FROM} invoices, pgLedger
	\texttt{WHERE} invoiceID = k \texttt{AND} pgLedger.user \texttt{IN} (S,
	M) \texttt{AND} pgLedger.commitTime > \texttt{now()} - \texttt{interval}
	`24 hours' \texttt{AND} invoices.xmax = pgLedger.txid \texttt{AND}
	invoices.deleter \texttt{IS NULL}; \\ \hline
	\end{tabular}
}
\end{table*}


\par
\textbf{Shared Memory Data Structures.} We introduced the following four data structures in
the shared memory (the last two are used only for the \textit{execute-order-in-parallel} flow).
\par
\begin{enumerate}
\item \texttt{TxMetadata} enables communication and synchronization between block processor and
backends executing the transaction (as needed in the commit phase). The block processor uses
this data structure to check whether all transactions have completed its execution and to signal
each backend to proceed further. Each entry consists of the global transaction identifier
(as present in the transaction), transaction ID assigned locally by the node, process id of the backend,
a semaphore to denote whether the transaction has completed its execution and waiting for
a signal to proceed further, and a final status (i.e., commit/abort).
\item \texttt{BlockProcessorMetadata} helps in signaling block processor from the backend once it
commits/aborts a transaction. Additionally, this
data structure holds the last committed block number, current block number (so that the backend
can use it to set the \texttt{creator} \& \texttt{deleter} block number in rows), and a
semaphore to enable signaling from middleware to block processor.
\item \texttt{TxWriteSet} holds the write-set of each transaction so that after SSI validation
and before writing to WAL, backend can store the \texttt{creator} or \texttt{deleter} in each row. 
Note, \texttt{TxWriteSet} just keeps a pointer to the row in buffer.
For an update, it stores both the old and new row pointer. For delete, it stores only the old row pointer.
Further, \texttt{TxWriteSet} is used to compute the hash of the write-set for each block.
\item \texttt{TxOrder} helps backend to apply our proposed \textit{block-aware
abort during commit} variant of SSI. It stores the global transaction identifier
of each transaction in the block as per the commit order 
to aid in finding whether a \texttt{nearCon-}
\texttt{flict} or \texttt{farConflict} is present in the same block.
\end{enumerate}

\textbf{Blockchain Related Catalog Tables.} We introduced two system catalog
tables, \textit{pgLedger} and \textit{pgCerts}. The \textit{pgLedger} is the \textit{ledger table} 
described in \S\ref{sec:recovery}, and stores information
about each transaction such as the global identifier, local transaction ID, query
passed in the transaction, client who submitted the transaction, and commit/abort status. 
It is used for recovery and for supporting provenance queries.
The \textit{pgCerts} table stores the cryptographic credentials of all blockchain users.
\par
\textbf{Provenance Query.} We introduced a special type of read only query called \textit{provenance}.
This query type can see all committed rows present in tables irrespective of whether it is
inactive (i.e., marked with \textit{xmax}) or active. As it can access all historical
content, it enables support for very complex analytical and
audit queries with the help of \textit{pgLedger}. Table~\ref{table:audit} presents
two examples of provenance queries.
\subsection{Components Modified} 
\label{sec:psql-modified-components}
\par
\textbf{Application Interface \& Deterministic PL/SQL Procedures.} We have enhanced the default application
interface, i.e., libpq, with additional APIs to submit blockchain transactions \& provenance
queries, and fetch the latest block height at the node. To make the PL/SQL procedure deterministic, we have
restricted the usage of date/time library, random functions from the
mathematics library, sequence manipulation functions, and system information functions.
Further, \texttt{SELECT} statements must specify \texttt{ORDER BY primary}
\texttt{\_key} when using
\texttt{LIMIT} or \texttt{FETCH}. Additionally, it cannot use row headers such as \textit{xmin},
\textit{xmax} in \texttt{WHERE clause}.
\par
\textbf{SSI Based on Block Height.} We have added two fields for each row: (i) \texttt{creator}
block number, (ii) \texttt{deleter} block number.
During commit, these two fields are filled for entries in the \texttt{TxWriteSet} depending on
whether an entry is an update, insert, or delete. SI applies a row visibility logic
using \textit{xmin} and \textit{xmax} to identify whether a row should be seen by a transaction
(as described in \S\ref{sec:psql-background}). We enhance the row visibility logic to have
additional conditions using the row's creator and deleter block number and the \texttt{snapshot-height} of the transaction
(as described in \S\ref{sec:execution-phase-eo}). This is in addition to the default row visibility logic 
that helps avoid seeing rows updated/deleted by concurrent transactions.
During predicate reads, the default visibility logic in
PostgreSQL traverses rows as per the index entries that satisfies the given predicate or
traverses the whole table when an index is missing. For our approach to work (mainly to avoid phantom or stale read described in
\S\ref{sec:execution-phase-eo}), all read access is enforced to happen via index entries only which satisfies
a given predicate clause. Otherwise, there is a high possibility of transactions getting aborted
due to false positive phantom or stale read. Hence, if an index is not available for the given
predicate clause, nodes abort the transaction. Further, \texttt{SELECT * FROM table\_name;}
is not allowed from PL/SQL procedures as it always traverses the whole table. It is possible
to implement our approach without the need for an index, but for simplicity we defer such optimizations.
Note, any kind of individual \texttt{SELECT} statements without any writes on the blockchain schema will not be
affected by SSI as the transaction would be marked as read-only and would be executed on one node only.
\par
\textbf{SSI Block-Aware Abort During Commit \& ww-dependency.} 
For the \textit{execute-order-in-parallel} approach, we have modified the abort rule to follow our proposed rules
in Table~\ref{table:abort-rule}. The modified SSI utilizes the \texttt{TxOrder}
data structure in the shared memory. For ww-conflicts, we allow writes to the same object by
different transactions (as updates are anyway not in-place in PostgreSQL) by maintaining an array
of \textit{xmax} values comprising of transaction IDs of all competing transactions in the row
being updated. During commit, for each old row entry in the \texttt{TxWriteSet}, the backend
(corresponding to the transaction that is committing now) checks \textit{xmax} values and marks all other
transactions for abort as only one transaction can write to the row to avoid lost update.
Finally, it retains only its own transaction ID in the \textit{xmax} field.

\subsection{Ordering Service}\label{sec:psql-ordering-service} As described in
\S\ref{sec:key-components}, any consensus algorithm can be leveraged.
For this work we use two different ordering services: an Apache
Kafka~\cite{kafka} and ZooKeeper~\cite{zookeeper} based crash fault tolerant, and
a BFT-SMaRt~\cite{bftsmart} based byzantine fault tolerant ordering service.
Clients/peers connect to independent orderer nodes to submit transactions and
receive created blocks. Orderer nodes either: 1.\ connect to a Kafka cluster and
publish all received transactions to a \textit{Kafka topic}, which delivers
the transactions in a FIFO order, or 2.\ submit the transactions to a BFT-SMaRt
cluster which is then responsible for totally ordering the received transactions
among them.
For creating a block of transactions, we use two
parameters: \textit{block size}, the maximum number of transactions in a block,
and \textit{block timeout}, the maximum time since the first
transaction to appear in a block was received. 
Each orderer node either publishes a \textit{time-to-cut} message to the
Kafka topic or sends it to the BFT-SMaRt cluster when its timer expires. The
first time-to-cut message is considered to cut a block and all other duplicates
are ignored.
Once a block is cut,
orderer nodes append their signatures to the block, persist the block in
filesystem and then send it to connected peers.
\subsection{Transaction Flow} \label{sec:psql-transaction-flow}
Figure~\ref{fig:trans-flow-execute-and-order-postgres} depicts the new and modified components described
in the previous sections and outlines the \textit{execute-order-in-parallel} transaction flow. Since the
\textit{order-then-execute} flow is simpler, we omit presenting the details for it in the interest of brevity.
The application leverages the blockchain interfaces in libpq to fetch the latest block height and submit
the transaction to the PostgreSQL backend. After verifying the client signature leveraging \textit{pgCerts}, 
the transaction is forwarded to other PostgreSQL nodes and to the ordering service using the communication middleware
(the client can also submit the transaction to all peers, rather than one peer forwarding to all other peers). 
The backend updates
\texttt{TxMetadata}, executes the transaction leveraging the SSI variant based on \textit{block-aware
abort during commit}, and collects the write-set into \texttt{TxWriteSet}. It then sets \textit{ready-to-proceed}
status in \texttt{TxMetadata} and waits for a signal from the block processor. 

Upon receiving the block from the ordering service, the middleware 
verifies the orderer signature and stores the block in \textit{pgBlockstore}. The block processor
retrieves each unprocessed block from the \textit{pgBlockstore} one at a time and adds all transactions to \textit{pgLedger}.
It confirms from \texttt{TxMetadata} that all transactions have completed execution and then serially signals each
backend as per the order in the block to proceed further. Upon receiving this
signal, the backend validates the transaction based
on the \textit{block-aware abort during commit}
logic as explained in \S\ref{sec:committing-phase-eo}, sets either \textit{commit}/\textit{abort} status in \texttt{TxMetadata},
and signals the block processor. Once all transactions have been processed, the block processor updates
the \textit{pgLedger} table with the status for each transaction.
We are yet to implement the checkpoint flow described in \S\ref{sec:checkpointing-phase-eo}.

\section{Evaluation}
\label{sec:experiments}
In this section, we study the performance and cost of both our design approaches.
We measure performance in terms of \textit{throughput} and \textit{latency}. \textit{Throughput} 
is defined as the average number of
unique transactions committed per second in the blockchain network, and \textit{latency}
is defined as the average time taken to commit a transaction, measured from the time
a client submits it. A transaction
is said to be committed in the blockchain network when majority of the nodes commit. The cost
is measured in terms of resource utilization such as CPU, memory,
disk and network bandwidth utilization.
We study the effect on performance and cost by varying several system parameters, namely
(a) block size, (b) transaction arrival rate, (c) smart contract query complexity, (d) deployment model
(local/wide area network), and (e) the number of database nodes (network size).
\par
We use three smart contracts (1) \texttt{simple contract}---inserts values
into a table; (2) \texttt{complex-join contract}---performs complex
joins between two tables to execute aggregate operations and write the result to
a third table; (3) \texttt{complex-group contract}---performs aggregate queries
over subgroups in a group and uses order by and limit to write the max aggregate
across the subgroups into a table. We illustrate these queries in
Appendix~\ref{apdx:smart contract}.
Note that while \texttt{simple contract} can be implemented in most blockchain platforms today, 
both \texttt{complex-join contract} and \texttt{complex-group contract} are impossible to implement efficiently. 
We consider two deployment models, the first where all organizations host their database nodes on a common
public cloud data center (LAN), and second where nodes are hosted independently
leading to a multi/hybrid-cloud setup (WAN). In our experiments, the WAN setup involved four data centers
spread across four continents. These models help us study the communication overheads of the two approaches.
In both deployments, database and orderer nodes were hosted on
virtual machines each with 32 vCPUs of Intel Xeon E5-2683 v3 @ 2.00GHz
and 64 GB of memory. In the multi-cloud deployment, the network bandwidth between nodes was between
50 and 60 Mbps, whereas, it is 5 Gbps in the single cloud deployment.
\par
In addition to \textit{throughput} and \textit{latency}, we also measure the following seven micro metrics
(all as averages) to gain a deeper understanding of system behavior:
\begin{enumerate}
	\item the block receive rate (\texttt{brr})---the number of blocks received per second at the
		middleware from orderer.
	\item the block processing rate (\texttt{bpr})---the number of blocks processed and committed per second
		at the block processor. 
	\item the block processing time (\texttt{bpt} in \textit{ms})---the time taken to process
		and commit a block.
	\item the block execution time (\texttt{bet} in \textit{ms})---the time taken to start all
		transactions in a block till they suspend for commit/abort. 
	\item the transaction execution time (\texttt{tet} in \textit{ms})---the time taken by the
		backend to execute a transaction until it suspends for commit/abort. 
	\item the block commit time (\texttt{bct} in \textit{ms})---the time taken to perform the
		serial commit of all transactions in a block and can be measured as $\texttt{bpt}-\texttt{bet}$.
	\item the missing transactions (\texttt{mt})---the number of transactions missing per second
		while processing blocks at the block processor (relevant for \textit{execute-order-in-parallel}). 
\end{enumerate}
We define two additional terms:
(1) the \textit{peak throughput} as the maximum achieved throughput for a given smart contract type
and block size across all possible arrival rates; (2) the \textit{system utilization} (\texttt{su}) as
the fraction of time the block processor is busy (\texttt{bpb}) expressed as a percentage. \texttt{bpb}
can be estimated as the average number of blocks processed in 1s (\texttt{bpr}) multiplied by the time
taken to process each block (\texttt{bpt}).
In an ideal scenario, when peak throughput is achieved, our system utilization should be close to $100\%$.
\par
Unless mentioned otherwise, our experiments use a single data center with three organizations, each running 
one database and one orderer node.
At the orderer, the block timeout was set to 1s and the block size was varied. We used \textit{pgTune}~\cite{pgtune} to configure
PostgreSQL and \textit{max\_connection} (i.e., number of backends) was set to 2600. The
arrival rate was load balanced among the orderer nodes in the case of \textit{order-then-execute} and the
database nodes in the case of \textit{execute-order-in-parallel}.
\begin{figure}
    \begin{center}
	    \includegraphics[scale=0.45]{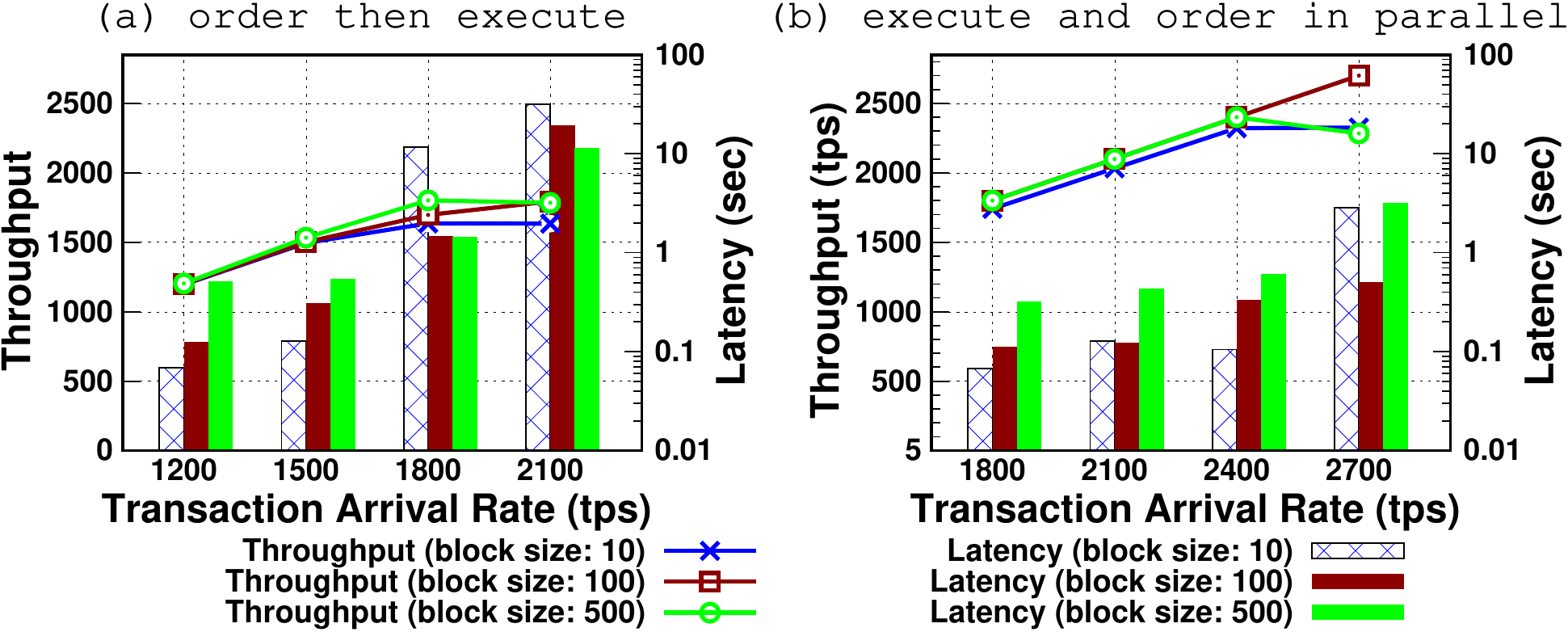}
  		\vspace{-.7cm}
		\caption{\small{Performance with \texttt{simple contract}}}
	    \label{fig:simple}
 \end{center}
  \vspace{-.6cm}
\end{figure}
\begin{table}
\small{
	\caption{\small{Order then execute: micro metrics for an arrival rate of 2100 tps.}}
	\label{table:oe-simple-micro}
	\begin{tabular}{  | p{.5cm} | p{.7cm} | p{.7cm} |  p{.7cm} | p{.6cm} | p{.6cm} | p{.5cm}| p{.7cm}|}
        \hline
	\texttt{bs} & \texttt{brr} & \texttt{bpr} & \texttt{bpt} & \texttt{bet}
	& \texttt{bct} & \texttt{tet} & \texttt{su} \\ \hline
	10 & 209.7 & 163.5 & 6 & 5 & 1 & 0.2 & 98.1\% \\ \hline
	100 & 20.9 & 17.9 & 55.4 & 47 & 8.3 & 0.2 & 99.1\% \\ \hline
	500 & 4.2 & 3.5 & 285.4 & 245 & 44.3 & 0.4 & 99.7\% \\ \hline
  \end{tabular}
	\vspace{-.4cm}
}
\end{table}

\subsection{Block Size and Arrival Rate}
\textbf{Order then Execute.}
Figure~\ref{fig:simple}(a) plots the \textit{throughput} and \textit{latency} achieved in
\textit{order-then-execute} approach using the \texttt{simple contract}.
Table~\ref{table:oe-simple-micro} presents the micro metrics for an arrival rate of 2100 transactions
per second (\textit{tps}).
With an increase in transaction arrival rate, the throughput increased linearly as expected till it
flattened out at around 1800 \textit{tps}, which was the peak throughput 
(system utilization close to $100\%$ as shown in Table~\ref{table:oe-simple-micro}).
When the arrival rate was close to or above the peak throughput, the latency increased significantly
from an order of 100s of milliseconds to 10s of seconds.
For an arrival rate lower than the peak throughput,
with an increase in the block size, the latency increased. The reason is that with an increase in block size,
it took longer for the orderer to fill a block with transactions leading to an increased block creation time,
and hence, a larger waiting time for each transaction at the orderer. For an arrival rate greater than the peak throughput,
with an increase in block size, the latency decreased. This is because there was no waiting for transactions to arrive
and form a block, and more transactions were executed in parallel. 
This is also observable as the block processing time (\texttt{bpt})
of a block of size $n$ was always lesser than the sum of \texttt{bpt} of $m$ blocks each of size $\frac{n}{m}$.
For the same reason, with an increase in the block size, the throughput also increased.
\par
\textbf{Execute and Order in Parallel.}
Figure~\ref{fig:simple}(b) plots the same for 
\textit{execute-and-order-in-parallel} approach. 
Table~\ref{table:eo-simple-micro} presents the micro metrics for an arrival rate of 2400 \textit{tps}.
The maximum throughput achieved was 2700 \textit{tps}, i.e., 1.5$\times$ higher than what was
achieved with \textit{order-then-execute}. Note that the system utilization (\texttt{su}) was
only 90\%. When the arrival rate was greater than 2700 tps, the throughput started to degrade and
\texttt{su} never reached 100\%. We believe the reason to be a large number of active backends
that resulted in a lot of contention on default shared memory datastructure (we intend to study this further 
as part of the future work). 
Though the block processing time (\texttt{bpt}) and the block execution time (\texttt{bet})
were observed to be lesser with \textit{execute-and-order-in-parallel} as compared to \textit{order-then-execute},
the
block commit time (\texttt{bct}) was observed to be higher. This could again be because of a
large number of active backends.
\par
\textbf{Comparison With Ethereum's Order then Execute.}
Blockchain platforms such as Ethereum also adopt an \textit{order-then-execute} approach, but execute transactions
serially once the block is formed. To emulate this behavior, we made our block processor also
execute and commit transactions one at a time. 
This resulted in a peak throughput of 800 \textit{tps} (for a block size of 100, although the block size 
does not matter when we execute serially).
This is only about 40\% of the throughput achieved with our approach, which supports parallel
execution of transactions leveraging SSI.
\begin{table}[htp]
\vspace{-.4cm}
\small{
\caption{\small{Execute and order in parallel: micro metrics for an arrival rate of 2400 tps.} }
\label{table:eo-simple-micro}
\begin{tabular}{  | p{.4cm} | p{.7cm} | p{.7cm} |  p{.7cm} | p{.6cm} | p{.5cm} | p{.45cm}|  p{.4cm}|p{.55cm}|}
\hline
	\texttt{bs} & \texttt{brr} & \texttt{bpr} & \texttt{bpt} & \texttt{bet}
	& \texttt{bct} & \texttt{tet} & \texttt{mt} & \texttt{su} \\ \hline
	10 & 232.26 & 232.26  & 3.86 & 2.05 & 1.81 & 0.58 & 479 & 89\% \\ \hline
	100 & 24.00 & 24.00 & 35.26 & 18.57 & 16.69 & 3.08 & 519 & 84\% \\ \hline
	500 & 4.83 & 4.83 & 149.64 & 50.83 & 98.81 & 6.27 & 230 & 72\% \\ \hline
\end{tabular}
\vspace{-.3cm}
}
\end{table}

\subsection{Smart Contract Complexity}
\begin{figure}
    \begin{center}
	    \includegraphics[scale=0.45]{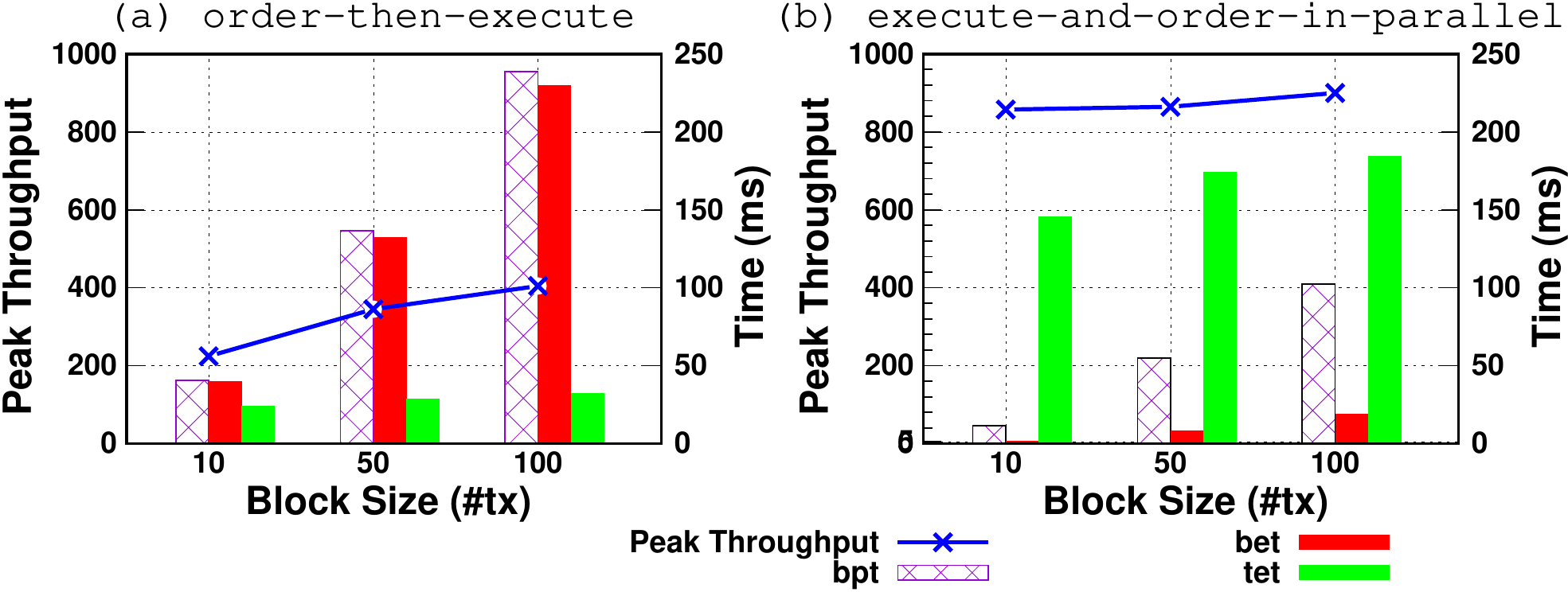}
  		\vspace{-.7cm}
		\caption{\small{Performance with \texttt{complex-join contract}}}
	    \label{fig:complex}
 \end{center}
  \vspace{-.2cm}
\end{figure}
\textbf{Order then Execute.}
Figure~\ref{fig:complex}(a) plots the peak throughput and micro metrics achieved when
using \texttt{complex-join contract} for \textit{order-then-execute}. As observed earlier, with an
increase in block size, the throughput increased and reached a peak throughput of 400 \textit{tps}. This was less than
25\% of what was observed with \texttt{simple contract}, primarily because of transaction execution time (\texttt{tet})
increasing by about 160$\times$ as compared to \texttt{simple contract}. The block processing time (\texttt{bpt}) and
the block execution time (\texttt{bet}) also increased.
The CPU and memory utilization for \texttt{simple contract} was 10\% and 15 GB respectively, compared
to 30\% and 15GB for \texttt{complex-join contract}.
\par
\textbf{Execute and Order in Parallel.}
Figure~\ref{fig:complex}(b) plots the same for \textit{execute-order-in-parallel}. 
Both the block processing time
(\texttt{bpt}) and the block execution time (\texttt{bet}) were lower as compared the one
observed in \textit{order-then-execute}. This is because, by the time the block reaches the
block processor, all transactions were either executing or already completed execution.
The peak throughput achieved 
was more than twice that of 
\textit{order-then-execute}.
Unlike \textit{order-then-execute} approach, this
approach permits concurrent execution of a larger number of transactions unrestricted by the block size
(in \textit{order-then-execute}, the maximum number of concurrently executing transactions is
capped by the block size). This manifested as a significantly larger increase to the transaction
execution time \texttt{tet} compared to \textit{order-then-execute}.
For the same reason, the CPU utilization increased to 100\%, i.e., 3.3$\times$ higher than what was observed for
\textit{order-then-execute}.
\par
\begin{figure}
    \begin{center}
	    \includegraphics[scale=0.45]{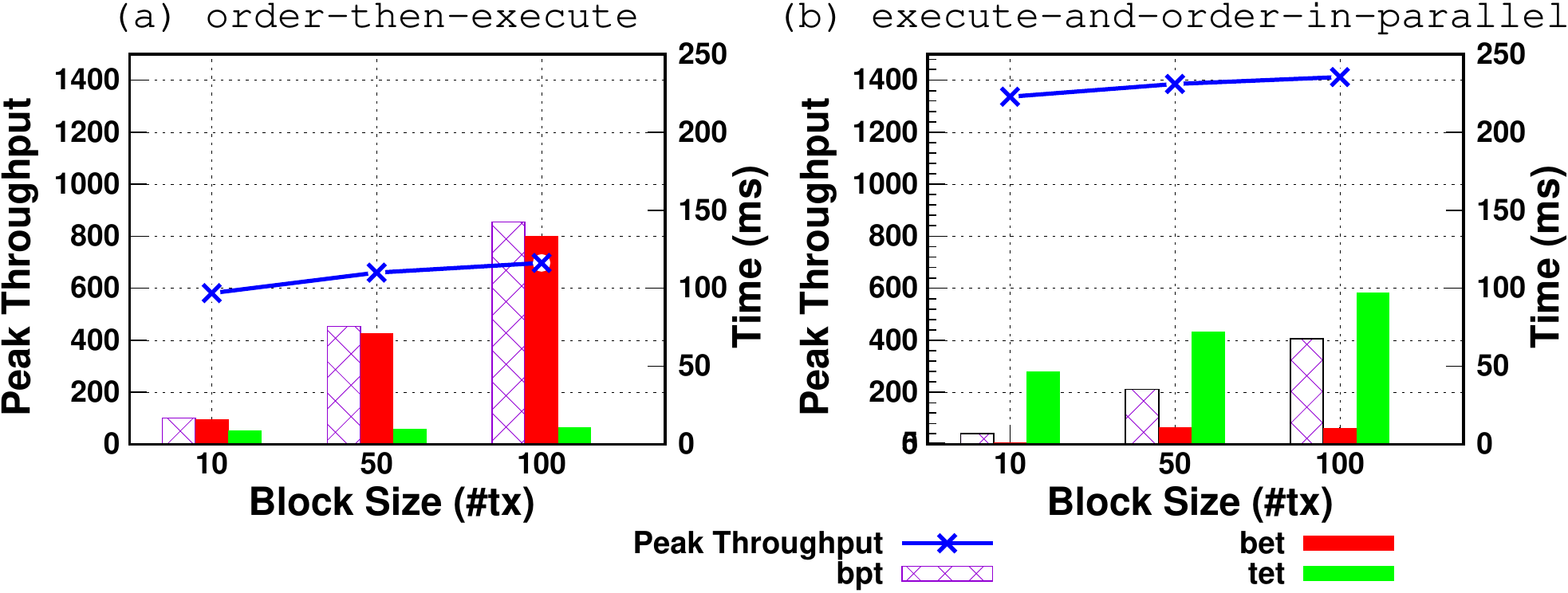}
  		\vspace{-.7cm}
		\caption{\small{Performance with \texttt{complex-group contract}}}
	    \label{fig:complex-group-perf}
 \end{center}
  \vspace{-.2cm}
\end{figure}

In the third \texttt{complex-group contract} smart-contract, for a block size of
100, the maximum throughput achieved for \textit{order-then-execute} and
\textit{execute-order-in-parallel} was 1.75$\times$ and 1.6$\times$ higher
respectively than what was observed with the \texttt{complex-join contract}.
We plot the graphs for the same in Figure~\ref{fig:complex-group-perf}.
\subsection{Deployment Model and Network Size}
\begin{figure}
    \begin{center}
	    \includegraphics[scale=0.45]{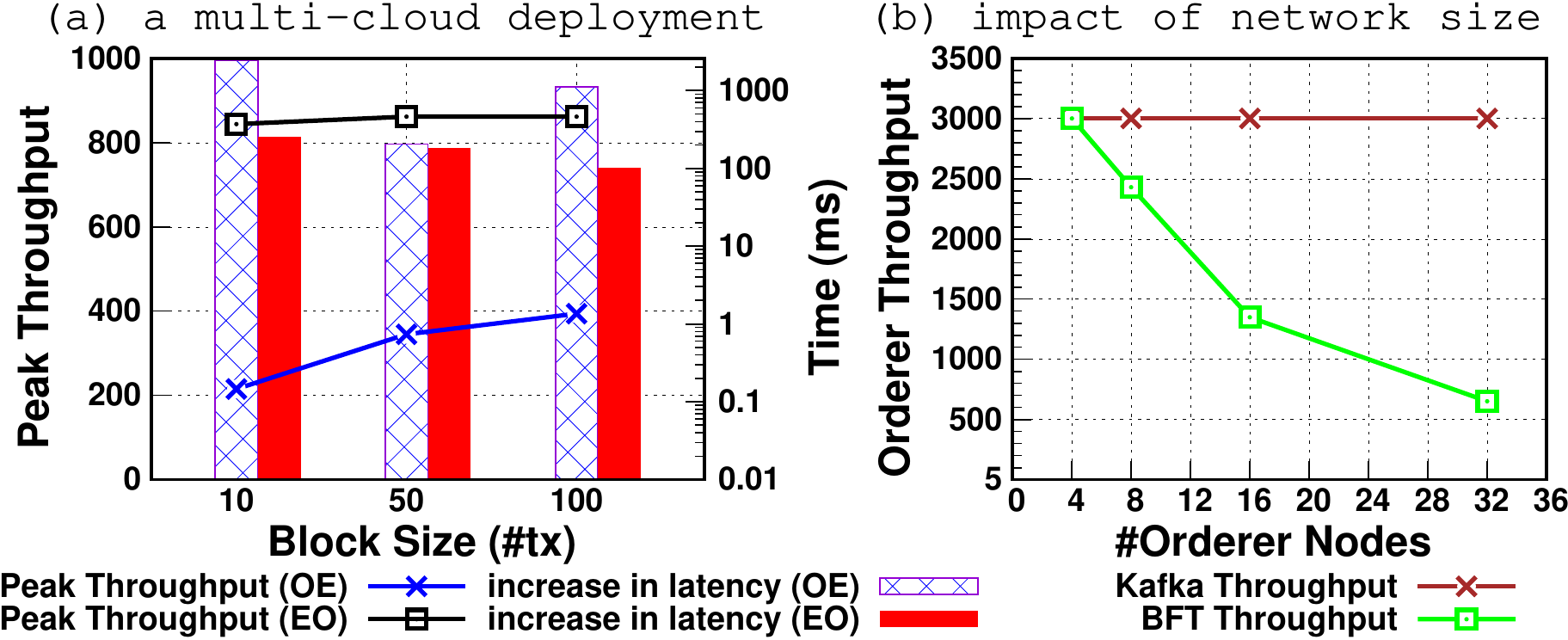}
  		\vspace{-.3cm}
		\caption{\small{Performance (a) withcomplex contractin amulti-cloud deployment;
						(b) with different network sizes.}}
		\label{fig:complex-wan-scale}
 \end{center}
  \vspace{-.6cm}
\end{figure}

\textbf{Deployment Model.}
Figure~\ref{fig:complex-wan-scale}(a) plots the peak throughput achieved with both approaches in a
multi-cloud deployment. As compared to a single cloud deployment (refer to Figure~\ref{fig:complex}),
only the latency increased on average by 100 \textit{ms} due to the WAN network but the throughput reminded
the same for most part except a 4\% reduction in the peak throughput when the block size was 100.
In our system, each transaction was only 196 bytes in size, and hence the size of a block with 500 transactions
was only about 100 KB. Hence, the lower bandwidth in the multi-cloud deployment did not have a significant impact on the
performance that our system can work equally well in a geographically distributed environment.
\par
\textbf{Network Size}
Figure~\ref{fig:complex-wan-scale}(b) plots the throughput achieved with kafka and bft based ordering service while
varying the number of orderer nodes and fixing the transaction arrival rate to 3000 tps. With
an increase in the number of orderer nodes, we observed no impact on the performance of kafka-based
ordering service but the performance achieved with bft-based ordering service reduced from 3000 tps
to 650 tps as expected due to the communication overhead. With an increase in the number of database
nodes alone, the overall system throughput did not get affected but limited by the peak throughput
of the ordering service.

\vspace{-.2cm}
\section{Related Work}
\label{sec:related}
\textbf{Bitcoin}~\cite{bitcoin} and \textbf{Ethereum}~\cite{ethereum}
adopt an order-execute model, where transactions are first ordered in a block through consensus
(such as proof-of-work) and each node validates and commits transactions to its copy of the blockchain
by serially executing each transaction. In our \textit{order-execute} approach, we leverage SSI 
to execute transactions concurrently. Further, such platforms only support
a simple key-value data model. 

\textbf{Hyperledger Fabric}~\cite{fabric} adopts an \textit{execute-then-order} approach, where transactions are first
executed and endorsed by multiple nodes, then ordered by consensus, and then followed by serial validation and commit.
In contrast, in our \textit{execute-order-in-parallel} approach transaction execution and ordering happen parallelly. 
Fabric only supports leveldb and couchdb as the underlying database, with support for composite keys and range queries. 
In contrast, we support the full spectrum of
SQL queries in a smart contract (for the first time) with the exception of libraries that could introduce non-determinism and
blind updates. Further, by leveraging capabilities already available within a relational database we provide support
for complex data types, schemas, constraints, and provenance queries. Performance studies on Fabric~\cite{fabric, perfFabric1}
have shown that a throughput of 3000 tps can be achieved with goLevelDB as the state database for a
\texttt{simple} smart-contract, while the throughput drops to 700 tps with CouchDB.
\par
\textbf{Hyperledger Composer}~\cite{composer} is a set of collaboration tools for building blockchain business networks utilizing
Hyperledger Fabric. It allows a user to write a basic SQL query with limited syntax~\cite{composer-query}
and internally converts the SQL query to a JSON selector query~\cite{couchdb-query} supported by CouchDB.

\textbf{Corda}~\cite{corda} refers to their platform as a decentralized database. 
Transactions are executed by one node at a time (not parallelly executed by all nodes) and the results are shared with other 
nodes that have a need to know. 
There is no notion of a blockchain in Corda. However, optionally, a notary
could order transactions and validate them for double spending. 
Corda uses the H2 embedded SQL engine as the state database. State objects can define a relational mapping, and an
object serialization framework is used to store the states in a relational database. This permits querying the
database using SQL and also permits rich queries (such as joins) with an organization's private non-blockchain data.
However, it does not enable rich query support within the smart contract itself.

\textbf{Veritas}~\cite{veritas} proposes shared verifiable tables 
using a set of Redis key-value stores each owned by an organization.
Only the verifiability property, i.e., immutable logs, is supported. For ensuring consistency across replicas,
it uses a centralized trusted timestamp server to order transactions. Further, a transaction
is executed only on one of the nodes, and each node periodically ships logs of multiple read-write sets
to other nodes via a Kafka-based broadcast service. Nodes are trusted to execute and share results
correctly with other nodes, and hence is not decentralized. Nodes vote on transactions to resolve
any conflicts.

\textbf{BigchainDB}~\cite{bigchaindb} employs Tendermint consensus~\cite{tendermint, tendermint-url} over a set
of independent MongoDB~\cite{mongodb} instances, 
each owned by a different organization. It supports immutability and decentralization. While the overall
goals of BigchainDB are similar to ours, there are fundamental architectural differences. It supports
only user tokens/assets similar to Bitcoin
and no support for smart contracts. A transaction submitted to BigchainDB is forwarded to
Tendermint nodes, which 
ensure that all the BigchainDB nodes
agree on the next block in a Byzantine fault tolerant way. 
Transactions are serially executed post ordering, similar to Ethereum.

It is evident from the related work that ours is the first blockchain relational database
with rich features of both blockchain and databases. Further, the \textit{execute-and-order-in-parallel}
approach proposed in this paper is highly novel and we believe it is applicable to
centralized replicated databases.

\section{Discussion and Future Work}
\label{sec:futurework}
In ongoing work, we are studying the performance of both our design
approaches in the presence of \texttt{rw}/\texttt{ww}
\texttt{dependencies} with two setups:
(a) a homogeneous network so that
all nodes would be approximately at the same \texttt{current-block};
(b) a heterogeneous network
that could lead to a significant gap in the \texttt{current-block}
of different nodes over time.
\par
As stated in \S\ref{sec:committing-phase-oe} and \S\ref{sec:recovery},
there is a potential to optimize the current implementation by (a) executing the
serialization check using all transaction's \textit{inConflictList}
and \textit{outConflictList} at once in a separate background
process instead of executing at the backend serially, and
(b) committing the whole block atomically rather than 
one transaction at a time. Optimizations are also possible in our communication
layer by leveraging gossip protocols (for larger number of nodes) and batching together 
transaction replication. We expect the performance to improve
with these optimizations.
\par
Over time, the amount of historical data in the network
would grow significantly. Hence, we need to enhance the existing
pruning tool such as \texttt{vacuum}~\cite{vacuum} to remove rows based
on fields such as \texttt{creator}, \texttt{deleter}.
As privacy is an important consideration for blockchain solutions, 
there is potential to enhance relational database
features such as user management 
to provide stronger privacy guarantees. Data privacy can be achieved using
database schemas and the concept of channels as in Hyperledger Fabric~\cite{fabric}.
\section{Conclusion}
\label{sec:conclusion}
In this paper, we presented the design of a \textit{blockchain relational database}, a decentralized
database with replicas managed by different organizations that do not trust one another. The key challenge we
addressed is in ensuring that all untrusted replicas commit transactions in the same serializable order that respects the
block ordering determined by consensus. 
We proposed two design
approaches that leveraged and modified SSI
to achieve this, and devised a new variant of SSI based on block height. Leveraging features
already available in databases enables us to better support complex data types,
schemas, complex queries and provenance queries that are not provided by blockchain platforms today. 
We implemented the system on PostgreSQL and presented detailed performance results.
\par

\bibliographystyle{abbrv}
\bibliography{ms}  
\onecolumn
\appendix
\section{Smart Contracts used for evaluation}
\label{apdx:smart contract}

		\begin{figure*}[h!]
		  \vspace{+.4cm}
		    \begin{center}
			    \includegraphics[scale=0.50, frame]{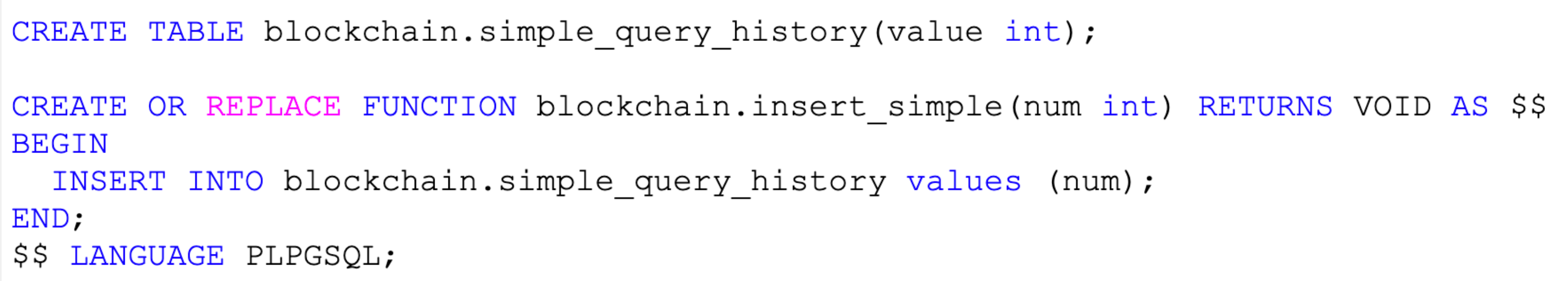}
				\vspace{-.3cm}
				\caption{Simple query with single inserts into table}
			    \label{fig:simple-contract}
		 \end{center}
		  \vspace{-.6cm}
		\end{figure*}
		\begin{figure*}[h!]
		    \begin{center}
			    \includegraphics[scale=0.48, frame]{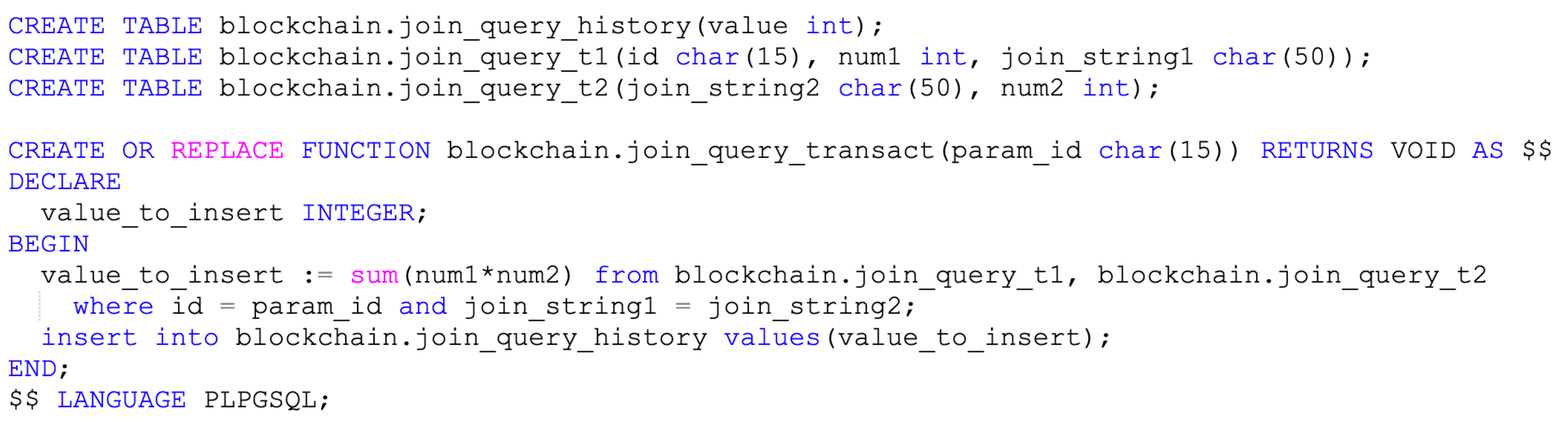}
				\vspace{-.7cm}
				\caption{Complex queries with joins and aggregates}
			    \label{fig:complex-join}
		 \end{center}
		  \vspace{-.6cm}
		\end{figure*}
		\begin{figure*}[h!]
		    \begin{center}
			    \includegraphics[scale=0.50, frame]{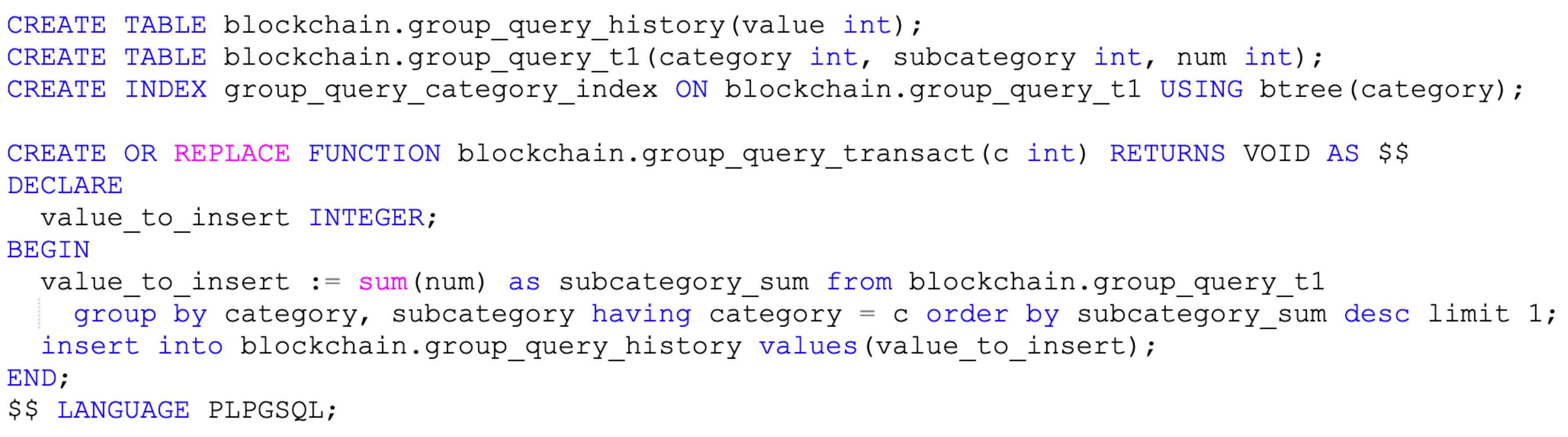}
				\vspace{-.7cm}
				\caption{Complex queries with group-by, order-by, limit and aggregates}
			    \label{fig:complex-group}
		 \end{center}
		\end{figure*}

\end{document}